\documentclass[smallheadings]{article}

\usepackage{tabularx,booktabs}
\newcolumntype{Y}{>{\centering\arraybackslash}X}

\usepackage{float}
\usepackage{longtable}

\usepackage[]{graphicx}
\usepackage{mathrsfs}
\usepackage{mdwlist}
\usepackage{amssymb}
\usepackage{eucal}
\usepackage{units}
\usepackage{rotating}

\usepackage{soul}
\usepackage[normalem]{ulem}

\usepackage{amsmath}
\usepackage{amsthm}
\usepackage{epstopdf}
\usepackage{dsfont}
\usepackage{bm}
\usepackage{enumerate}
\usepackage[usenames, table]{xcolor}
\usepackage{hhline}
\definecolor{blue}{rgb}{0,0,1}
\definecolor{purple}{rgb}{0.6,0,1}

\usepackage{natbib}
\usepackage{multirow} 
\usepackage[margin=1.2in]{geometry}
\usepackage{pdfpages}
\usepackage{float}
\usepackage{setspace}
\usepackage{afterpage}

\usepackage{pdflscape}
\usepackage{marvosym}

\usepackage{pgfplots}
\pgfplotsset{compat=1.18} % or another version installed on your system

\usepackage{adjustbox}

\usepackage{minitoc}
\usepackage[toc,page,header]{appendix}

\usepackage{multibib}
\newcites{appendix}{References}

\newcommand{\indep}{\perp \!\!\! \perp}

\newcommand\citeapos[1]{\citeauthor{#1}'s (\citeyear{#1})}

\usepackage{tikz}
\usetikzlibrary{positioning, arrows.meta, tikzmark,calc, matrix}

\usepackage[colorlinks = true, citecolor = blue, linkcolor = blue, urlcolor  = blue,]{hyperref}

\usepackage{rotating}
\usepackage[labelfont=bf]{caption}

\newcommand{\insertbottomrule}{\bottomrule}

\usepackage{array}
\newcolumntype{L}[1]{>{\raggedright\let\newline\\\arraybackslash\hspace{0pt}}m{#1}}
\newcolumntype{C}[1]{>{\centering\let\newline\\\arraybackslash\hspace{0pt}}m{#1}}
\newcolumntype{R}[1]{>{\raggedleft\let\newline\\\arraybackslash\hspace{0pt}}m{#1}}
\newcolumntype{Y}{>{\centering\arraybackslash}X}

\newcommand{\bmath}{\begin{eqnarray}}
\newcommand{\emath}{\end{eqnarray}}
\newcommand{\bmathnn}{\begin{eqnarray*}}
\newcommand{\emathnn}{\end{eqnarray*}}
\newcommand{\bi}{\begin{itemize}}
\newcommand{\ei}{\end{itemize}}

\renewcommand{\d}{\textrm{do}}

\usepackage{ bbold }

\theoremstyle{plain}

\theoremstyle{plain}

\DeclareMathOperator{\cov}{cov}
\usepackage{setspace}
\begin{document}

\title{Mental Models of Causal Structure\\ in Economics and Psychology}
\author{Sandro Ambuehl, Rahul Bhui, Heidi C. Thysen \footnote{Ambuehl: Department of Economics and UBS Center for Economics in Society, University of Zurich, Bl\"uemlisalpstrasse 10, 8006 Z\"urich, Switzerland, sandro.ambuehl@econ.uzh.ch. Bhui: MIT Sloan School of Management, 100 Main Street, Cambridge, MA 02142, USA, rbhui@mit.edu. Thysen: Department of Economics, Norwegian School of Economics, Helleveien 30, 5045 Bergen, Norway, heidi.thysen@nhh.no.}}

\maketitle
\thispagestyle{empty}

\begin{abstract}
A burgeoning literature in economics studies how people form beliefs about the causal structures linking economic variables, and what happens when those beliefs are mistaken. We survey this research and connect it to a rich literature in cognitive science. After providing an accessible introduction to causal Directed Acyclic Graphs, the dominant modeling approach, we review theory and evidence addressing three nested questions: how individuals reason within a fully parameterized causal structure, how they estimate its parameters, and how they learn such structures to begin with. We then discuss methodological challenges and review applications in microeconomics, macroeconomics, political economy, and business.\\ 
\textit{JEL-codes: D01, D03, D83, D84} 
\end{abstract}

\setcounter{tocdepth}{3}
\setcounter{secnumdepth}{3}

\newpage
\tableofcontents
\newpage

\section{Introduction}
\setcounter{page}{1}

Despite the familiar adage, facts rarely speak for themselves. Economic agents interpret the world through subjective mental models that govern what they learn and how they act.
A worker who attributes their low bonus to insufficient effort rather than their manager's hostility will fail to search for a better job. A voter who sees interest rates and prices rising in lockstep blames the central bank for the very inflation it is trying to fight. A firm that advertises heavily every December credits the campaign for holiday sales that might have happened anyway. Thus, even when the facts are not in question, their interpretation may be.

In this paper, we review the burgeoning body of research in economics on subjective mental models of causal structure, and synthesize it with the vast foundational literature from cognitive science.

Traditional research on belief updating in economics typically assumes that individuals know the structure of their environment and asks whether they update beliefs by the correct magnitude. Yet, the structure itself is uncertain in many economically relevant settings. Rather than operating with a perfect understanding, people must rely on simplified mental models, often incomplete or misspecified, to map observations into beliefs and actions. Economists have therefore paid increasing attention to subjective causal models to better bridge the gap between the objective environment and the behavior of its inhabitants. A growing collection of work explores the implications across areas as diverse as political economy, contract theory, behavioral economics, macroeconomics, marketing, and strategic management. This work speaks to phenomena such as cycles of populism, strategic persuasion, systematic cognitive biases, demand for counterproductive macroeconomic policies, misjudgments of product efficacy, and entrepreneurial innovation.

Beginning decades earlier, an extensive parallel literature in cognitive science has investigated how humans learn, represent, and reason about causal structure. One of the enduring puzzles in this tradition is how humans can draw far-reaching inferences from relatively sparse and fragmentary experience---that is, how we appear to `learn so much from so little'. Rather than learning isolated associations, people often seem to build structured causal representations that constrain interpretation and support generalization. Structure learning, in this view, is a central mechanism by which cognition achieves both efficiency and flexibility.

Despite the shared concerns of economics and cognitive science in this domain, the interaction between the two has been remarkably limited. Cross-field citations are rare, and collaborations rarer still. This disconnect is striking because both fields rely on the same formal framework of \emph{causal Directed Acyclic Graphs} (\emph{DAGs}) developed in computer science and artificial intelligence \citep{Pearl.2009, KollerFriedman}. This shared language offers a powerful opportunity for cross-fertilization. For example, cognitive science provides rich empirical evidence on how humans conceive of causal structure, while economics offers sharp theoretical tools for analyzing the behavioral and equilibrium consequences of such conceptions.

This review takes a step toward connecting the fields. We introduce economists to insights from cognitive science, while highlighting open areas that are particularly salient from an economic perspective. For instance, the most prominent economic applications of the framework consider agents who interpret data through misspecified mental models. These applications have remained largely theoretical because the empirical literature in economics on the topic is still nascent, while computational cognitive science has focused more on how people learn the actual structure of the world than on the behavioral consequences of misspecification.

Economists may object that much evidence shows individuals struggle with Bayesian updating even in relatively simple environments \citep{Benjamin.Review}, raising doubts about their ability to reason in settings with unknown causal structure. This concern rests on a particular benchmark for success, namely precise calibration of numerical probability judgments within a well-specified model. But the work reviewed here concerns a much broader set of questions about how agents represent and infer causal structure---which variables they treat as related, in which direction, and through which pathways---and how they use that structure to interpret evidence, predict the effects of interventions, and evaluate competing narratives. The causal perspective helps consolidate a number of stylized facts about belief updating that are not naturally captured by models that treat beliefs as unstructured associations between variables. For example, learning that an outcome has one plausible cause reduces the weight placed on alternative causes (\emph{explaining away}); conditioning on an intermediate variable attenuates inferences about more distant causes (\emph{path blocking}); inferences based on passive observation differ systematically from those based on active intervention (\emph{observation--intervention distinction}); and people readily imagine situations that have never been and never will be observed (\emph{counterfactual thinking}). Indeed, subpar performance in some canonical reasoning problems may reflect the lack of an intuitively coherent causal structure. For instance, when the standard mammography problem is augmented with an explicit causal account of false positives (benign cysts), people are less prone to base-rate neglect \citep{KrynskiTenenbaum}.

Causal thinking comes naturally enough that even young children display its hallmarks \citep{gopnik2001causal, GopnikEtAl2004}. Nor is it confined to high-level cognition; Bayesian causal inference unifies many findings about more elementary forms of human information processing, including sensory perception and motor control \citep{ShamsBeierholm}. In fact, \citet{RichensEveritt} prove that any (human or artificial) agent capable of generalizing well across a family of counterfactual data generating processes (DGPs) must have learned, at least implicitly, a causal model of the relevant variables. Hence, causal models are not merely optional refinements to probabilistic reasoning, but are often necessary for approximately optimal decision-making.

The empirical evidence we review is predominantly experimental. Experiments sometimes raise external validity concerns, both regarding stake size and because they typically take place within the very brief time period of a single experimental session. Nevertheless, several studies in experimental economics show that biases in human judgment and reasoning do not disappear (though they are sometimes slightly attenuated) when stakes are raised by orders of magnitude \citep{CamererHogarth1999,enke2023cognitive}. Moreover, research that extends laboratory studies in psychology over timespans of several days or weeks shows that the qualitative results of the single-session studies often generalize \citep{WillettRottman2021AccuracyLongTimeframes,zhang2024causal,RottmanZhang2025LearningChangeOverTime}.

The topics we review extend far beyond the scope of an article-length treatment. We largely limit our coverage of the economic theory literature to learning with misspecified subjective models that take the form of causal DAGs \citep[see also][]{Spiegler.Review}. Focusing on a different set of questions, \citet{BohrenHauser.Review} review the literature on more general forms of misspecified learning.
Our coverage of formal methods is correspondingly brief. \cite{Pearl.2009} provides a comprehensive treatment of causal DAGs from a statistical point of view, to which \cite{Pearl.BookOfWhy} is an accessible introduction. \cite{KollerFriedman} present the broader formal framework of probabilistic graphical models used to operationalize many of these ideas.
The cognitive science literature on causal cognition includes many book-length treatments. To mention but a few, \cite{SlomanBook} offers a gentle introduction, \cite{waldmann2017oxford} reviews causal cognition more widely, and \cite{BayesianModelsOfCognitionBook} comprehensively survey Bayesian models of cognition.

\begin{figure}[htb]
\caption{The three levels of causal cognition.}
\label{fig:threeLevels}
\centering
\includegraphics[width=0.85\textwidth]{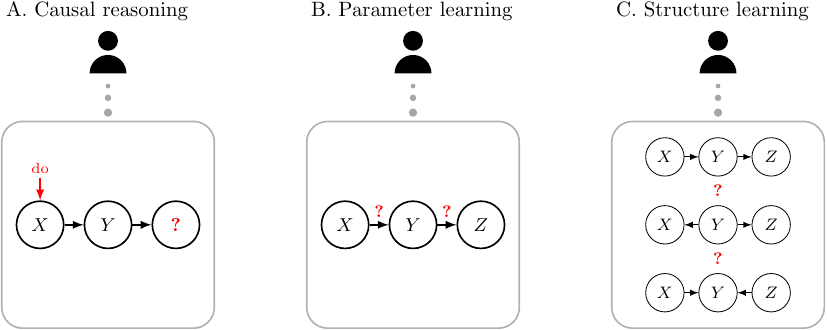}
\end{figure}

The body of this review starts with an informal introduction to causal DAGs (Section \ref{sec:dagsIntro}).
Three nested questions then organize the article, as illustrated in Figure \ref{fig:threeLevels}. First, how do individuals reason within a given causal model, assuming such a model is in place and its parameters are known (\textit{causal reasoning}; Section \ref{sec:causalReasoning})? Second, how do individuals learn about the magnitude of causal effects within a given structure, and what systematic errors do they make when their beliefs about structure are wrong (\textit{parameter learning}; Section \ref{sec:magnitudeLearning})? Third, how do individuals form beliefs about causal structure in the first place (\textit{structure learning}; Section \ref{sec:structureLearning})? Each of these sections first presents the relevant theoretical foundations---how idealized learners would draw these inferences---followed by empirical evidence that highlights the regularities, the inconsistencies, and the unanswered questions in each domain. We continue by discussing methodological advances and challenges in measuring beliefs about structure (Section \ref{sec:measurement}). Section \ref{sec:applications} surveys applications of causal cognition across various subfields of economics. The concluding Section \ref{sec:conclusion} highlights several questions beyond the scope of this review.

\section{An informal introduction to causal DAGs}\label{sec:dagsIntro}

The workhorse framework in the literature on causal cognition is the causal DAG. This model encodes possibly stochastic causal relationships among variables of interest. It consists of a set of nodes, each representing a random variable, and directed edges that represent direct causal influence, such as in Panel C of Table \ref{tab:dagExample}. An arrow from $X$ to $Y$ indicates that intervening on $X$ may change the distribution of $Y$, holding all else fixed. Much of the framework's power derives from its ability to represent causal structure in the absence of any functional form assumptions. Fully specifying the distributional assumptions renders the causal DAG a \emph{Causal Bayesian Network}. The framework was originally developed in the 1980s by Judea Pearl in statistics and computer science (see \cite{Pearl.2009} for a textbook treatment), introduced to economics by \cite{Spiegler.QJE}, and taken up in a growing psychology literature around the turn of the millennium, including \citet{tenenbaumGriffiths2001}, \citet{glymour2001mind}, \citet{GopnikEtAl2004}, \citet{SlomanBook}, and \citet{lagnado2007beyond} among others, building on precursors such as \citet{waldmann1992predictive}, \citet{waldmann1995causal}, and \citet{plach1999bayesian}.

Economists unfamiliar with causal DAGs often find it easiest to understand them through the example of systems of regression equations. Consider Panel A of Table \ref{tab:dagExample}. It involves standard normal random variables $\epsilon_i$ and real-valued parameters $\beta_{ij}$ for $i, j \in \{X, Y, Z\}$. These quantities define three random variables $X, Y, Z$. Importantly, we imbue this system with causal meaning by interpreting each equation as how the left-hand-side variable changes if we intervene on the right-hand-side variable. In this way, $X$ is defined as exogenous (since it does not depend on any other variable) while $Y$ and $Z$ are endogenous (since they do depend on other variables).

Panel C of Table \ref{tab:dagExample} shows the DAG representing this system. Each left-hand side variable is a node. To draw the arrows, we consider each variable that appears on the left-hand side of an equation, starting, for instance, with $Z$. Variable $Y$ appears on the right-hand side of that equation, indicating that it exerts a direct causal influence on $Z$. To represent this fact in the DAG, we draw an arrow from $Y$ to $Z$. No other variable appears on the right-hand side of that equation, so we do not draw any further arrows into $Z$. Next, we consider the equation with $Y$ on the left-hand side. Variable $X$ appears on the right-hand side. Accordingly, we draw an arrow from $X$ into $Y$. Finally, the equation for $X$ has no variables on the right-hand side. Hence, no arrows point into $X$. Because the system of equations is recursive, the resulting graph is acyclic. Endowing the DAG with a causal interpretation, which means reading arrows as direct causal mechanisms that determine the effects of intervention, renders it a \emph{causal DAG}.

The literature uses evocative terms to describe properties of nodes in a network: $Y$ is called the \textit{parent} of $Z$, and $Z$ the \emph{child} of $Y$. $X$ is a called \textit{root} or \emph{ancestral} node since it has no parents. $Z$ has no children, so $Z$ is a \textit{sink} or \textit{terminal} node. The \textit{ancestors} of $Z$ are $X$ and $Y$, the variables that cause it directly or indirectly; the \textit{descendants} of $X$ are the $Y$ and $Z$, the variables that it causes directly or indirectly.

\begin{table}
\caption{Representations of example probabilistic causal models. \label{tab:dagExample}}
\begin{center}
\includegraphics[width=0.55\textwidth]{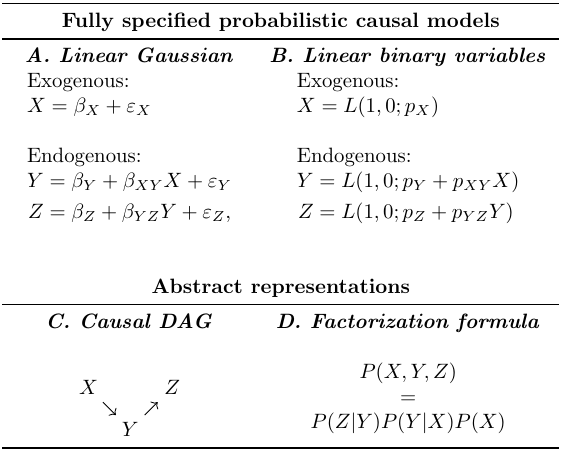}
\end{center}
\footnotesize
\textit{\textbf{Notes:}} We use $L(a, b; q)$ to denote the lottery that pays $a$ with probability $q$ and $b$ otherwise. The factorization formula represents only the probabilistic information embedded in the causal DAG (conditional independences), but not the causal information.
\end{table}

To illustrate the level of abstraction at which causal DAGs operate, Panel B of Table \ref{tab:dagExample} presents a second example with binary random variables, a setting that frequently appears in the economic theory literature \citep[e.g.,][]{EliazSpiegler.Narratives}. Here, $X$ is an exogenous indicator variable, and $Y$ and $Z$ are binary outcomes whose probabilities of success depend on their parents. $L(a, b; q)$ denotes the binary lottery that equals $a$ with probability $q$ and $b$ otherwise. While the success probability of $X$ is constant, that of $Y$ and $Z$ depend on the realizations of other variables, creating dependence between the nodes. We derive the DAG representing this system as before. For instance, because $Y$ appears on the right-hand side of the equation that determines the distribution of $Z$, we draw an arrow from $Y$ to $Z$. Despite these differences between the systems in Panels A and B, their DAG representations are identical. The two systems share the same causal structure in the sense of which variables influence which variables, even though they differ substantially in their probabilistic details.

A common question is whether the causal DAG framework is restrictive because it allows causality between variables to flow in one direction only, whereas many models in economics as fundamental as the Keynesian cross involve feedback loops. The standard resolution is to accommodate feedback loops by unfolding the system over time: rather than representing a cycle between, say, income and consumption, one indexes each variable by period, so that income in period $t$  affects consumption in $t+1$, which in turn affects income in $t+1$ and so on. The resulting graph is acyclic. Such unfolding makes the model dynamic, however. Much of the pertinent literature considers static settings. These can incorporate feedback loops if one replaces the part of the structural system that involves feedback loops with its reduced form, which expresses the endogenous variables as functions of the exogenous variables alone.

\paragraph{Causation vs. correlation}

Economists are intimately familiar with the conceptual distinction between correlation and causation (sometimes interpreted as prediction and intervention, respectively). The framework of causal DAGs encompasses both.

We represent correlational statements using \emph{conditional probabilities}, written in the form $P(X|Y)$; these capture the \emph{flow of information}. Consider, for instance, the second equation of the causal system in Panel A of Table \ref{tab:dagExample}, represented by the link $X \rightarrow Y$ in Panel C. Algebraically inverting this equation yields $X = \frac{1}{\beta_{XY}}(Y - \beta_Y - \epsilon_Y)$, which highlights that variation in $Y$ partly reflects variation in $X$. From our assumption on the causal structure, however, we know that any such information cannot represent a causal effect of $Y$ on $X$.

To calculate causal effects, we use \emph{interventional probabilities}, written in the form $P(X|\d(Y))$. Such an expression represents the distribution of $X$ if $Y$ is exogenously set to a specific value. To see the difference to conditional statements, consider our example DAG. There, $Y$ does not have a causal effect on $X$, which we write as $X|do(Y) = X$. $Y$ does, however, have a causal effect on $Z$. From the structural equation we see that if we set $Y$ to a specific value $y$, we obtain the distribution of $Z$ after this intervention as $Z|\d(Y=y) = \beta_Z + \beta_{YZ} y + \varepsilon_Z$. This example illustrates the general mechanics of the $do$-operator, which \citet{pearl2012} calls \emph{graph surgery}: when we intervene on a variable (that is, when we apply the $do$-operator to that variable), we replace the original DAG with a copy that disconnects that variable from its usual causes and replaces it with a constant. The distribution of variables in this altered model correctly captures the post-interventional distribution of all variables. The causal effect on $Z$ from setting $Y$ to 1 rather than 0, for instance, is then given as $P(Z|\d(Y=1))-P(Z|\d(Y=0))$. The formal rules of the $do$-operator are detailed in \citet{pearl2012}. They are consistent with the well-developed intuition about causal interventions that economists typically possess.\footnote{There are some cases where economists' common intuitions might be misleading or specific to a narrow class of models. One example concerns the intuition that total effect $=$ direct effect $+$ indirect effects which is true only in linear models.}

Many tools of modern applied microeconomics (such as natural experiments and regression discontinuities) can all be thought of as applications of the $do$-operator in real-world data. In \cite{angrist1990lifetime}, for instance, the draft lottery serves as a natural experiment to identify the causal effect of serving in the military. The lottery disconnects the event of serving in the military from its usual causes (such as socioeconomic background) that determine military membership in non-war times and acts as an intervention to determine membership.

\paragraph{Calculations with causal DAGs}
To perform calculations with causal DAGs, we represent the probabilistic information encoded in the DAG with the \emph{Bayesian network factorization formula}. It relies on the fact that we can write any joint probability distribution as a product of conditional probabilities, in the following way. Consider some $n$ random variables $X_1, ..., X_n$. By the definition of conditional probability, $P(X_1,X_2, ..., X_n)$ $= P(X_n|X_1, ..., X_{n-1})P(X_1, ..., X_{n-1})$. If we apply this definition repeatedly, we obtain the \emph{chain rule of probability}:\footnote{The symbol $P$ in the foregoing expression represents a probability measure. That is, $P$ is evaluated on sets of values that the collection of random variables $(X_1,X_2, \ldots, X_n)$ can take.
When we are interested in specific realizations $x_1, ..., x_n$, we read $P(X_1,X_2, \ldots, X_n)$ as $P(\{X_1=x_1\},\{X_2=x_2\}, \ldots, \{X_n=x_n\})$. Following the canonical convention that uppercase symbols denote random variables and lower case letters specific realizations, the foregoing expression is often written as $P(x_1, x_2, \ldots, x_n)$.}
\bmath\label{chainRuleGeneral}
P(X_1,X_2, \ldots, X_n) =  P(X_n|X_1, \ldots, X_{n-1})\cdot P(X_{n-1}|X_1, \ldots, X_{n-2}) \cdot \ldots \cdot  P(X_2|X_1)P(X_1).
\emath
In our three-equation system of Panel B, we can thus write 
\bmath\label{chainRuleExample}
P(X, Y, Z)=P(Z|X, Y)\cdot P(Y|X)\cdot P(X)
\emath
The causal structure represented in the DAG allows us to simplify this formula. The crucial information in the system concerns the  variables that are \emph{excluded} from the right-hand side in any given equation. For instance, while $Z$ could, in principle, depend on every other variable, the third equation in the system specifies that $Z$ depends only on $Y$. Therefore, once we know $Y$, the value of $X$ does not provide any additional information about the distribution of $Z$. In other words, the third equation guarantees that $Z$ is independent of $X$ conditionally on $Y$. Formally, $P(Z=z|Y, X) = P(Z=z|Y)$. From the second equation in the system, we know that the distribution of $Y$ depends on $X$, so that the distribution $P(Y|X)$ cannot be simplified further. Using these insights in equation (\ref{chainRuleExample}), we obtain\footnote{When we apply the chain rule of equation (\ref{chainRuleGeneral}), we order the variables. We directly obtain the DAG's factorization into parent-conditional probabilities only if we order the variables such that whenever $X_i\rightarrow X_j$ in the DAG, then $X_i$ comes before $X_j$. If we chose a different order, we would obtain, for instance, $P(X, Y, Z) = P(X)\cdot P(Z|X)\cdot P(Y|X,Z)$, where none of the factors can be simplified using the conditional independencies of the DAG. Moreover, note that the factorization formula is simply a way of describing the joint distribution over all variables, and hence does not retain the causal information embedded in a causal DAG.}
\bmath\label{factorizationFormulaExample}
 P(X, Y, Z) = P(Z|Y)\cdot P(Y|X)\cdot P(X).
\emath
While we can derive quantitative predictions using the machinery of systems of linear regression equations when all variables are jointly normally distributed, the factorization formula allows us to do this no matter the functional form of the variables' distributions.\footnote{Applying it correctly requires appropriately summing over all causal paths by which the intervened-on variable can change the other variables. For example, if we seek to calculate the effect on $Z$ of increasing $X$ from 0 to 1 in the system of Panel B of Table \ref{tab:dagExample}, we need to marginalize over $Y$, that is, to sum over all possible values that $Y$ could take, weighted by their respective probability: $P(Z=1 \mid do(X=x)) = \sum_{y\in\{0,1\}} P(Z=1\mid Y=y, X=x)\cdot P(Y=y\mid X=x)$. The factorization formula allows us to evaluate this as $P(Z=1\mid Y=1)\,P(Y=1\mid X=x) + P(Z=1\mid Y=0)\,P(Y=0\mid X=x) = (p_Z + p_{YZ})(p_Y + p_{XY}x) + p_Z(1 - p_Y - p_{XY}x) = p_Z + p_{YZ}(p_Y + p_{XY}x)$. The causal effect of $X$ on $Z$ is the difference of this expression evaluated at $x=1$ and $x=0$, which equals $p_{XY}\cdot p_{YZ}$, which is the product of the effects along the chain.} The statistical literature contains additional formula that aid in calculating causal effects in specific cases, such as the \emph{front-door} and \emph{back-door adjustment} formulas to help with confounding variables, and the \emph{mediation formula} to decompose total effects into direct and indirect parts \citep{Pearl.2009}.

\section{Causal reasoning} \label{sec:tokenLearning}\label{sec:wellStudiedProperties}\label{sec:archetypicalStructures}\label{sec:causalReasoning}

We turn first to causal reasoning: How should idealized learners draw inferences within a given DAG when they know both its structure and the strength of each causal link? And what do humans actually do? We focus on the three archetypal causal structures depicted in Figure \ref{fig:archetypalDAGs}). They illustrate key phenomena in causal reasoning and have been studied extensively in the empirical literature \citep[][provide a comprehensive review]{RottmanHastie.Review}. The structure $C \rightarrow M \rightarrow E$ is called a \emph{chain}, where $C$, $M$, and $E$ are mnemonics for $C$ause, $M$ediator, and $E$ffect, respectively. The structure $C_1\rightarrow  E  \leftarrow C_2$, in which an effect depends on two causes, is called a \emph{$v$-collider} or a \emph{common effect} DAG.  In the structure $E_1 \leftarrow C \rightarrow E_2$, a single cause influences two effects. It is called a \emph{fork} or \emph{common cause} DAG. We first use these three structures to illustrate three fundamental theoretical concepts: the Causal Markov Condition (how DAGs encode conditional independencies), collider bias (how conditioning on a common effect makes causes dependent), and Markov equivalence (how some DAGs yield identical observational implications). We then outline the empirical evidence concerning causal reasoning.

\subsection{Theoretical concepts}

\begin{figure}[htb!]
\caption{Archetypal causal structures \label{fig:archetypalDAGs}}
\begin{center}
\begin{tabular}{p{0.2\textwidth}
                >{\centering\arraybackslash}m{0.2\textwidth}
                >{\centering\arraybackslash}m{0.2\textwidth}
                >{\centering\arraybackslash}m{0.2\textwidth}}
& (A) & (B) & (C) \\ 
\toprule
Common labels & Chain, & Common Effect, & Common Cause, \\ & Line & $v$-collider & Fork \\[10pt]
&
    \begin{tikzpicture}[main/.style={}, baseline=(current bounding box.north)]  
        \node[main] (C) at (0,0) {$C$};  
        \node[main] (M) at (0.75,-1.2) {$M$};  
        \node[main] (E) at (1.5,0) {$E$};  
        \draw[->] (C) -- (M);
        \draw[->] (M) -- (E);
    \end{tikzpicture}
&    
    \begin{tikzpicture}[main/.style={}, baseline=(current bounding box.north)]  
        \node[main] (C1) at (0,0) {$C_1$};  
        \node[main] (C2) at (1.5,0) {$C_2$};  
        \node[main] (E) at (0.75,-1.2) {$E$};  
        \draw[->] (C1) -- (E);
        \draw[->] (C2) -- (E);
    \end{tikzpicture}
&
    \begin{tikzpicture}[main/.style={}, baseline=(current bounding box.north)]  
        \node[main] (C) at (0.75,0) {$C$};  
        \node[main] (E1) at (0,-1.2) {$E_1$};  
        \node[main] (E2) at (1.5,-1.2) {$E_2$};  
        \draw[->] (C) -- (E1);
        \draw[->] (C) -- (E2);
    \end{tikzpicture}
\\
\\
Example 
&
Salt consumption increases blood pressure which decreases life expectancy
&
Wealth and ability both determine admission to a selective university
&
Holiday seasonality drives both high sales and more advertising \\
\bottomrule
\end{tabular}
\end{center}
\end{figure}

\paragraph{Chain structure} 

The three-node chain (Panel A of Figure \ref{fig:archetypalDAGs}) illustrates a central principle of causal reasoning known as the \emph{Causal Markov Condition}. In the chain, the cause $C$ can only indirectly affect the effect $E$, through the mediator $M$. Hence, when holding the mediator fixed, changes in $C$ can no longer influence $E$; the mediator \emph{screens off} the cause from the effect. In correlational terms, the chain structure implies that $\cov(C, E | M) = 0$.\footnote{Formally, the structure implies conditional independence $C\indep E | M$, which is stronger than correlation and is defined even if the variables are multinomial rather than real-valued. Because many economists are more familiar with thinking about about covariance than about independence, we use correlational statements when doing so creates no confusion.} The Causal Markov Condition states generally that, conditional on all its direct causes (i.e., parents), a variable is independent of all variables that are not its downstream effects (i.e., descendants). Simply put, once the values of a variable’s direct causes are known, information about how those causes themselves came about adds nothing further. A deeper consequence is that, while correlation does not imply causation, it does carry some information about causal structure. For example, if we observe a system in which $\cov(C, E | M) \neq 0$, we can infer that the DGP cannot be a chain. More broadly, any DAG implies a list of conditional independence relationships, each stating that one set of variables is independent of another, conditional on a third (possibly empty) set. Comparing these implications to the conditional independence relationships present in the data allows the analyst to narrow down the candidate set of DAGs that may have generated the data.

\paragraph{Common effect structure}

\begin{figure}
\caption{Collider bias / Berkson's paradox}
\label{fig:colliderBias}

\begin{center}
\includegraphics[width=0.9\textwidth]{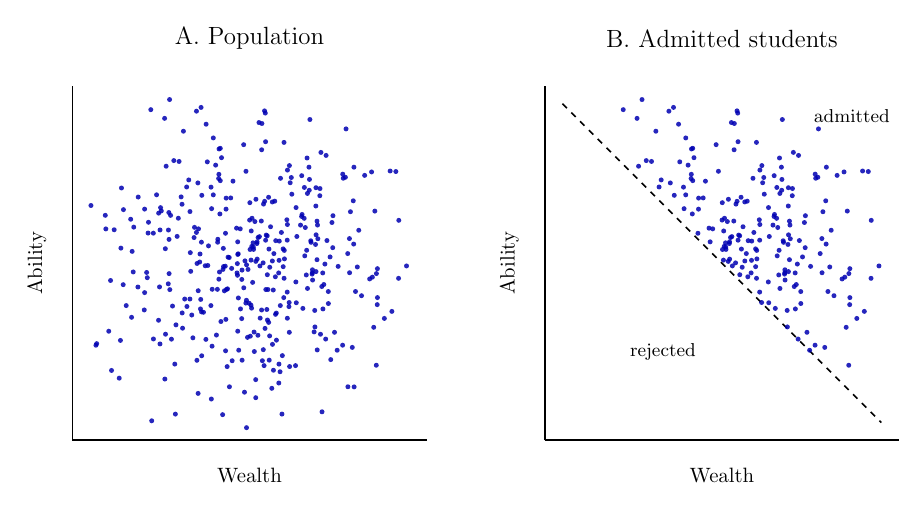}
\end{center}
\footnotesize
\textit{\textbf{Notes}}: Even if ability and wealth are uncorrelated in the population, they become correlated once we condition on the event of being admitted to a selective university that selects students based on both ability and wealth.

\end{figure}

Some of the most distinctive implications of causal inference in DAGs arise in common effect structures (Panel B of Figure \ref{fig:archetypalDAGs}), where multiple causes influence a single effect (such an effect is called \emph{collider node}). In these settings, conditioning can have the opposite effect from the screening-off property discussed above. Rather than blocking associations, conditioning on a common effect \emph{induces} correlations between its causes---a phenomenon known as \emph{collider bias}. Formally, a common effect structure implies that, generically, $\cov(C_1, C_2) \neq \cov(C_1, C_2 | E)$ (if $C_1$ and $C_2$ are real-valued). While this result may be surprising, it matches people's intuitive predictions. Suppose for example that academic ability and parental wealth are uncorrelated in the population, so that information about one is generally uninformative about the other. Consider now the student body at a university that admits students either on ability or through parental donations. Once we learn that a student has wealthy parents, we become more pessimistic about that student’s ability compared to other students at the university, since parental wealth may have compensated for lower ability in the admissions decision. As Figure \ref{fig:colliderBias} illustrates, conditioning on the effect ($E$, having been admitted to the university) creates a correlation between the two otherwise unrelated variables, ability ($C_1$) and parental wealth ($C_2$). If, as in the admissions example, collider bias arises from selection, it is known as \emph{Berkson's paradox}. Collider bias also explains other apparent paradoxes, such as the celebrated Monty Hall Problem, in which the door Monty opens depends both on the door the contestant selected and the location of the car.\footnote{In the Monty Hall problem a contestant in a game show chooses one of three doors, behind one of which is a prize (typically a car) and behind the other two are goats. After the contestant's choice, the host, Monty Hall, who knows what is behind each door, opens one of the remaining doors to reveal a goat, then offers the contestant the chance to switch to the other unopened door. The counterintuitive result is that switching is the optimal strategy and doubles the probability of winning from 1/3 to 2/3. As \citet{Pearl.BookOfWhy} explain, conditioning on the door Monty opened creates a (non-causal) correlation between the two determinants of his choice; it is this correlation that renders switching the optimal strategy.}

The admissions example illustrates the phenomenon of \emph{explaining away} (sometimes referred to as \emph{causal discounting}), a special case of collider bias which has been studied extensively in the psychology literature \citep{khemlani2011one}. In common effect structures with binary variables and independent or substitutable causes (e.g., where either cause alone can generate the effect), observing that one cause was present reduces the posterior probability that another cause was also present. Intuitively, when one sufficient explanation for an outcome is identified, alternative explanations become less likely. Formally, explaining away occurs when $P(C_1=1|E=1) > P(C_1=1|E=1, C_2=1)$. This inequality arises when one cause matters less for producing the effect once the other cause is present (in the sense that the likelihood ratio $P(E=1|C_1=1, C_2) / P(E=1|C_1=0, C_2)$ is decreasing in $C_2$).\footnote{To see this, note that by Bayes' rule and independence of the two causes, $P(C_1=1\mid C_2=c_2,E=1) = \frac{p_{1 c_2}\,\pi_1}{p_{1 c_2}\,\pi_1 + p_{0 c_2}(1-\pi_1)}$, where $p_{c_1 c_2} = P(E=1\mid C_1=c_1,C_2=c_2)$ and $\pi_1 = P(C_1=1)$. This expression is increasing in the likelihood ratio $p_{1 c_2}/p_{0 c_2}$, so the decreasing likelihood ratio condition $p_{11}/p_{01} < p_{10}/p_{00}$ directly implies $P(C_1=1\mid C_2=1,E=1) < P(C_1=1\mid C_2=0,E=1)$, which in turn implies $P(C_1=1\mid C_2=1,E=1) < P(C_1=1\mid E=1)$ by the law of total probability.}
Substitutability of the causes is critical; when causes are complementary (for instance, when both are necessary to cause the effect) learning that one cause was present can make it more rather than less likely that the other cause was also present.

Collider bias occurs regardless of whether the multiple causes of an effect are independent (as in Panel B) or whether they can causally affect each other (as would be the case, for instance, if the DAG featured an additional edge $C_1 \to C_2$). Yet, common effect structures in which the causes of an effect are \textit{not} directly causally linked---a class called called \emph{$v$-colliders} (where the letter $v$ is a graphical mnemonic)---have fundamentally different implications when fitting models to data than colliders whose parent nodes are linked (Section \ref{sec:parameterLearning}). DAGs in which no collider is a $v$-collider are called \emph{perfect} (`all parents are married').

\paragraph{Common cause structures}

In common cause structures (Panel C of Figure \ref{fig:archetypalDAGs}), a single cause leads to two separate effects. The Causal Markov Condition therefore also applies in these networks: holding fixed the common cause, the realization of one effect provides no information about the other effect. Formally, $E_1$ and $E_2$ are independent conditional on $C$. 

Importantly, the correlational implications of the common cause structure coincide with those of the chain $E_1 \to C \to E_2$, for which the Causal Markov Condition also implies that $E_1$ and $E_2$ are independent conditional on $C$. This is the only independence relationship in either structure.

This example illustrates the core concept of \emph{Markov equivalence}. While distinct DAGs always differ in their full set of \emph{interventional} implications, they sometimes make identical \emph{conditional predictions}---that is, they sometimes entail the same set of statistical associations in observational data. Such DAGs are called Markov equivalent. An immediate implication for structure learning (covered in Section \ref{sec:structureLearning}) is that correlational information alone cannot distinguish between Markov-equivalent DAGs. Relatedly, when making conditional predictions based on passive observation it does not matter whether one has fit a misspecified DAG to the data as long as that DAG is Markov equivalent to DAG that characterizes the DGP.

Determining Markov equivalence may appear hard because deriving the full set of correlational implications for a DAG is generally laborious. Fortunately, \citet{VermaPearl1990} provide  a straightforward criterion: two DAGs are Markov equivalent if and only if they have the same \emph{skeleton} and the same set of $v$-colliders (called \emph{$v$-structure}). One obtains the skeleton of a DAG by dropping all arrowheads, that is, by making the directed graph into an undirected one. Using this criterion, we can quickly see that the chain (panel A of Figure \ref{fig:archetypalDAGs}) and the common cause structure (panel C) are Markov equivalent: both have an empty $v$-structure, and once we drop the arrowheads, they look the same. We can also see that they are not Markov equivalent to the common effect structure (Panel B) which features a non-empty $v$-structure. Hence, an analyst considering these three candidate structures can, based on observational data alone, infer whether or not the DGP is a common effect structure, but cannot distinguish whether it is a chain or a common cause structure. 

Common cause structures play an important role in cognitive science because they give rise to the phenomenon of \emph{cue combination}, a well-studied topic in domains as fundamental as sensory perception \citep{trommershauser2011sensory}. It concerns how multiple signals are integrated, based on their reliability, to infer an underlying latent cause, which entails computing $P(C|E_1, E_2)$. This logic is familiar to economists. Multiple noisy observations (e.g., financial indicators, expert assessments, or survey responses) can each be informative of an underlying state, reflecting the aggregation of conditionally independent evidence about a shared cause.

\subsection{Empirical evidence}\label{sec:evidenceCausalReasoning}

Causal reasoning has been studied extensively in cognitive science, typically using externally provided causal structures wrapped in cover stories. Some studies take inspiration from everyday physical, social, or biological systems, where participants’ judgments may naturally recruit existing domain knowledge. When seeking to limit the influence of priors, studies deliberately feature unfamiliar or fictional settings, such as interactions between novel chemicals or science-fiction vignettes about aliens. Studies also vary by whether they verbally describe the probabilities that parametrize the DAGs, present data about individual realizations in tabular form, or have subjects experience a sequence of realizations (called a trial-by-trial design).

Overall, experimental research finds that subjects’ behavior is broadly consistent with normative predictions, in the sense that `when the normative calculations imply that an inference should increase, judgments usually go up; when calculations imply a decrease, judgments usually go down' \citep{RottmanHastie.Review}. One important caveat, however, is that subjects’ judgments tend to vary less with information and experimental manipulations than predicted. This attenuation is reminiscent of a general tendency for conservative belief updating, possibly for the same reasons that underlie conservative updating in balls-and-urns tasks \citep{Benjamin.Review, behavioralAttenuation}. Studies in which subjects learn probabilities by experience tend to find greater adherence to the normative benchmark than those that describe probabilities verbally. Beyond these general patterns, the literature also documents various deviations from predictions in specific causal structures.

\paragraph{Causal Markov Condition}

Do experimental subjects' judgments adhere to the Causal Markov Condition? In many cases, people clearly recognize its qualitative implications and reason accordingly. In \citet{AmbuehlThysen}, for instance, subjects could increase their payoff by selecting which of two causal models was consistent with correlational data generated by an underlying DGP, based on verbal descriptions using terms such as `symptom', `cause', and `mediator' as well as graphical depictions. Around two thirds of subjects consistently identified the correct model when doing so required intuiting that fixing the mediator in a chain DAG blocks the influence of the cause on the effect. Relatedly, \citet{SussmanOppenheimer2011} show that when people predict an unknown variable from two observed variables in a deterministic setting, the weight they place on each predictor tracks the conditional independence structure implied by the Causal Markov Condition. In chain and common cause structures, their subjects placed nearly all weight on the screening-off variable and effectively ignored the other variable, even though they weighted them roughly equally when both predictors were independent causes of the target.

These studies consider settings that are either deterministic \citep{SussmanOppenheimer2011} or feature large amounts of data that eliminate uncertainty about the magnitudes of correlations \citep{AmbuehlThysen}. They also present subjects with abstract variables that do not suggest any real-world analogs.
However, when the underlying causal networks feature more stochasticity, when subjects are presented with descriptions of causal relations between real-world variables, and when they are asked to internalize these structures rather than choose between them, humans reliably violate the Causal Markov Condition, as \citet{RottmanHastie.Review} conclude. \citet{Rehder2014}, for instance, asked subjects to assume that three variables were linked in a chain DAG. Even when he described the value of the middle node as fixed, subjects judged the effect as being more likely to occur when the upstream cause (the ancestral node in the chain) was present rather than absent, in violation of the screening off property. Similarly, with a common cause structure, subjects used one observed effect to draw inferences about the other effect even when the cause was observed. In a common effect structure with independent causes, subjects tended to treat the causes as correlated despite no apparent reason for this. These patterns do not appear to be driven by time pressure, suggesting that they reflect systematic features of causal reasoning \citep{Rehder2014, KolvoortEtAl2024}, and they persisted when causal relations were described between abstract variables rather than real-world entities \citep{Rehder2014}. In line with a more general tendency \citep{LejarragaHertwig}, some studies in which subjects learn DAG parameters by experience rather than from description find greater adherence to the normative benchmark \citep{RehderWaldmann2017}, though evidence is mixed \citep{RottmanHastie}; see Section \ref{sec:parameterLearningEmpirics}.

Several explanations for Markov violations have been proposed \citep{RottmanHastie}. One possibility is that subjects posit an unobserved causal mechanism linking variables that are assumed to be independent in the experimenter's model, effectively enriching the causal structure beyond what is explicitly described \citep{rehder2005feature, ParkSloman2013, marchant2023context}. Consider, for instance, the example $fitness \gets jogging \to weight\;loss$. Taking this DAG at face value, conditioning on $jogging$ eliminates the correlation between $fitness$ and $weight\;loss$. In the real world, however, jogging is not the only physical activity that can influence these variables; biking, say, has a similar effect. Subjects asked to judge the correlation between $fitness$ and $weight\;loss$ holding constant $jogging$ may therefore draw on this richer real-world model, in which variables such as $biking$ generate a residual correlation. Such subjects will appear to violate screening off, even though their judgments are perfectly consistent with a more complete causal structure.\footnote{In one experiment of \citet{ParkSloman2013} for instance subjects reason about two consequences of living in Africa ($C$). In one treatment the two effects are both mediated by the same mechanism: subjects are told that the intense sunlight causes people to wear sunglasses ($E_1$) as well as long sleeves ($E_2$). In the other treatment, long sleeves are framed as malaria protection, and hence are mediated through a different mechanism. Across multiple such examples, they find that in the same-mechanism condition, $P(E_1 \mid C)$ drops from 75.8 to 66.5 when subjects additionally learn that $E_2$ is absent, in violation of screening-off. In the different-mechanism condition, by contrast, the corresponding drop is a much smaller; from 75.1 to 72.0.} Relatedly, subjects may believe that the observed variables exhibit measurement error and are imperfect indicators of underlying states, in which case the conditional independencies implied by the stated DAG may not apply in their subjective model \citep{rehder2005feature}. A different account emphasizes associative biases whereby reasoners treat variables as more likely to co-occur when they share causal links, even when such associations should be screened off given the structure \citep{rehder2017failures}. Yet another explanation focuses on the cognitive process, proposing that people compute approximate inferences by mentally sampling states of a causal network in a biased way \citep{davis2020process, kolvoort2023bayesian}. Nevertheless, we are not aware of any account that is broadly considered able to accommodate all empirical patterns reported in the literature.

\paragraph{Explaining away} 

Building on the seminal empirical study of \citet{jones1967attribution}, \citet{MorrisLarrick} asked subjects to read an essay supporting Fidel Castro and informed them that there were two possible reasons why it might have been written rather than an essay criticizing him. With a 50\% chance, the author was free to choose which view to express; otherwise, the author was instructed to write either a pro-Castro or an anti-Castro essay, with the two assignments being equally likely. Once subjects were informed that the author was in the random-assignment condition, their belief that the author personally favored Castro dropped significantly. The assignment condition thus explained away the hypothesis that the author held pro-Castro views.

With deterministic causal structures, even young children exhibit explaining away \citep{GopnikEtAl2004}. In the classic `blicket detector' task, children observe that certain objects (`blickets') activate a machine when placed on it, while others do not. Multiple objects are initially presented together, leaving open which of them is causally responsible. When children later learn that one object alone is sufficient to activate the machine, they correspondingly reduce their belief that a second object is also a cause.

In the literature that has followed \citep[see][]{RottmanHastie.Review, khemlani2011one}, explaining away occurs often but not always, and when it does, it is typically attenuated relative to the Bayesian benchmark. Judgments about the probability that one cause occurred generally do move in the predicted direction when information about an alternative cause is revealed, but less strongly than normative models predict. The strength of explaining away varies substantially across individuals and experimental contexts \citep{SussmanOppenheimer2011}, and is sensitive to task features. Like failures of screening off, failures of explaining away seem more common when probabilities are learned from description than from experience \citep{RehderWaldmann2017}. Failures of explaining away do not, however, seem to be related to time pressure \citep{KolvoortEtAl2024}. Various theories have been proposed to describe these empirical patterns \citep{khemlani2011one}, including those which address Markov violations \citep{RottmanHastie}, such as \emph{causal elaboration} \citep[in which reasoners consider or invent causes outside the explicit problem description][]{rehder2005feature, ParkSloman2013, marchant2023context} and biased sampling \citep{davis2020process, kolvoort2023bayesian}.

\paragraph{Markov equivalence}

Markov-equivalent DAGs produce identical correlational predictions but differ in their implications for intervention. For such structures, causal direction should therefore be irrelevant when making predictions from observed associations, but should matter when choosing actions. Do people adhere to these principles?

A body of evidence answers this question broadly in the affirmative \citep{SlomanHagmayer2006, HagmayerFernbach2017}. Experiment 4 of \citet{HagmayerSloman}, for instance, used two Markov-equivalent structures, a chain and a common cause over the same three variables. The action served as root cause in the chain and as one of the two effects in the common cause structure; participants were queried about the presence of an auxiliary variable that functioned as mediator in the former case and as common cause in the latter. When the action was merely observed, participants' inferences about the auxiliary variable were identical across structures, consistent with the principle that Markov-equivalent DAGs are indistinguishable from observational data. However, when the action was deliberately chosen or externally forced, they considered the auxiliary variable more likely to be present in the chain condition, as prescribed by the \emph{do}-operator.\footnote{Closely related experiments in \citet{HagmayerSloman} and \citet{HagmayerMeder2013} find similar results in settings where the underlying structures are not strictly Markov equivalent (as they do not contain exactly the same variables) but are observationally equivalent in a broader sense (because some variables are unobserved).}

However, research on the \emph{predictive/diagnostic reasoning asymmetry} has documented a systematic deviation. By Markov equivalence, the two DAGs $X\to Y$ and $X \gets Y$ have the same correlational properties. Yet people tend to be more confident in their predictive inferences from cause to effect than in their diagnostic inferences about causes from effects. \citet{TverskyKahneman.CausalSchemas} first documented this effect by asking subjects whether it is more likely that a blue-eyed mother's daughter also has blue eyes, or that a blue eyed girl's mother also has blue eyes. Although both conditional probabilities should be equal, the overwhelming majority believed the former to be more likely. Similarly, they believed that predicting a son's height from his father's height would be more accurate than predicting a father's height from his son's height. People thus behave as if they perceive greater predictive strength along the direction of a causal arrow (\emph{predictive reasoning}) than in the reverse direction (\emph{diagnostic reasoning}). In some cases, the effect appears to reflect a tendency to consider alternative causes (e.g., the father's eye color in the blue-eyes example) when reasoning from effect to cause, but to neglect them when reasoning from cause to effect. This difference in the consideration of alternatives is sometimes taken to define the asymmetry \citep{FernbachDarlowSloman1,FernbachDarlowSloman2}.

\section{Parameter learning}\label{sec:magnitudeLearning}\label{sec:parameterLearning}

In the previous section, we considered agents who reason within a fully parametrized DAG. What if an agent only knows the causal structure, represented by the DAG, but not the magnitudes or strengths of its causal links? These magnitudes can be inferred based on the available data, paralleling standard statistical approaches. We outline the theoretical concepts before reviewing the relevant empirical evidence (Section \ref{sec:parameterLearningEmpirics}).

\subsection{Theoretical concepts}

We first discuss parameter learning when the DAG is correctly specified (Section \ref{sec:magnitudeLearningRational}). We then turn to learning with misspecified models (Section \ref{sec:misspecifiedLearning}), which is more closely tied to the kinds of non-standard behaviors studied in behavioral economics.

\subsubsection{Parameter learning with correctly specified models}\label{sec:magnitudeLearningRational}

\paragraph{Fitting a DAG to data}

If a reasoner has in mind a DAG that describes the causal relations between variables, how can she use data to learn about the strength of these relations? When the variables in a DAG are discrete, doing so is conceptually straightforward. The DAG factorization formula lets us express the joint distribution of all variables as a product of conditional probabilities. Replacing these conditional probabilities with their sample counterparts (the observed proportions in the data) yields the maximum likelihood estimates of the parameters of the corresponding Bayesian network, and thus of the joint distribution it implies \citep[][Chapter 17]{KollerFriedman}. This process is called `fitting the DAG to the data,'  'parameterizing the DAG,' or `projecting the data through the DAG.' 

\definecolor{hlyellow}{RGB}{255,250,205}
\begin{table}[htb]
\caption{Dataset for DAG fitting}
\label{tab:contingencyTable}
\begin{center}
\includegraphics[width=0.4\textwidth]{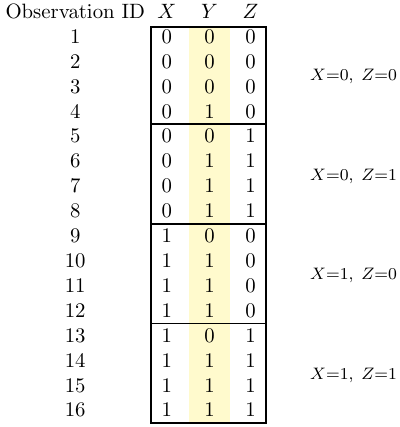}
\end{center}
\footnotesize
\textbf{\textit{Notes}}: Hypothetical dataset with $N=16$ observations of three binary variables $X,Y,Z$.
Each rectangle groups the four observations sharing a given combination of $X$ and $Z$.
The highlighted column marks $Y$, the dependent variable
in the collider DAG $X\rightarrow Y \leftarrow Z$. Reading off the fraction of $Y=1$
within each rectangle yields $P(Y|X,Z)$.
\end{table}

To illustrate, suppose we wish to fit the collider DAG $X\rightarrow Y \leftarrow Z$ to the data in Table \ref{tab:contingencyTable}. Its factorization formula is $P(X,Y,Z)=P(Y|X,Z)P(X)P(Z)$. We can estimate $P(X)$ from the dataset by counting all observations for which $X=1$, yielding $P(X=1)= 0.50$. In a similar way, we obtain $P(Z=1)=0.50$. We read off the collider node $Y$'s conditional distribution by focusing on all observations with a given combination of $X$ and $Z$ (that is, within each rectangle in Table~\ref{tab:contingencyTable}) and calculating the fraction of those observations for which $Y=1$. For example, the top rectangle contains the four observations with $X=0$ and $Z=0$, one of which has $Y=1$, yielding $P(Y=1|X=0,Z=0)=0.25$. Repeating this calculation for each of the four rectangles yields the full conditional probability table for $Y$. In the present example, this reveals that $P(Y=1|X,Z)=0.75$ if $X=1$ or $Z=1$, and 0.25 when both $X=0$ and $Z=0$. This result suggests that the function that determines the state of $Y$ depending on $X$ and $Z$ (called the \emph{link function} or \emph{connector}) is similar to a logical OR function: $Y$ is `on' with high probability if either input node is `on,' and with low probability otherwise.\footnote{A closely related formal concept, the \emph{noisy-OR} specification, models each cause as having an independent probability of producing the effect. By implication, if both causes are absent, the effect is absent with certainty. The \emph{leaky noisy-OR} specification generalizes the noisy-OR by allowing the effect to occur with some probability even if no cause is present.}

Observe that this inference requires no functional form assumptions; the OR-like relationship is revealed directly by the data. Had the data instead shown that $P(Y=1|X,Z)$ is high only when both $X$ and $Z$ are 1 but low otherwise, we would instead have inferred an AND-like relationship.

While the dataset in Table \ref{tab:contingencyTable} was indeed generated from the collider $X\rightarrow Y \leftarrow Z$, we can also fit the chain DAG  $X\rightarrow Y \rightarrow Z$ to this data. This DAG is misspecified given the DGP, but the exercise illustrates the mechanics of parameter fitting for different causal structures. The factorization formula for the chain DAG is $P(X,Y,Z)=P(Z|Y)P(Y|X)P(X)$. As before, $P(X = 1) = 0.50$. Because $Z$ is not a parent of $Y$ in this DAG, we estimate $P(Y=1|X)$ by \emph{marginalizing} over $Z$, meaning that we sum over all observations with a given value $X$ regardless of their value of $Z$. We find that $P(Y=1|X=1)=0.75$ and $P(Y=1|X=0)=0.50$. Similarly, we obtain $P(Z=1|Y=1)=0.60$ and $P(Z=1|Y=0)=0.33$. As will become relevant later, we see that interpreting the data in Table \ref{tab:contingencyTable} through the lens of the chain suggests that raising $X$ increases the probability that $Y=1$, because $P(Y=1|X=1)>P(Y=1|X=0)$, which in turn increases the probability that $Z=1$, because $P(Z=1|Y=1)>P(Z=1|Y=0)$. Hence, viewing this dataset through the lens of a chain DAG (wrongly, in this case) suggests that $X$ can exert an indirect causal effect on $Z$.

This estimation approach can require a large amount of data. If one has additional information, such as knowledge of functional form, more statistically efficient approaches may be used. An analyst facing continuous variables that are jointly Gaussian, for instance, can estimate parameters with linear regression.\footnote{The joint Gaussian assumption implies that there are no interaction terms in the conditional mean, because if $X$ is an $n$-dimensional vector of jointly Gaussian random variables, then $E(X_i | X_J)$ is linear in $X_J$ for all $i\in\{1,\ldots,n\}$ and $J\subseteq\{1, \ldots, n\}$.}

\paragraph{Bayesian learning}

The above section describes maximum likelihood estimation of DAG parameters. The approach can be naturally extended to a Bayesian framework. Doing so yields not only point estimates but entire posterior distributions over parameters. This information is needed for Bayesian structure learning (Section \ref{sec:structureLearning}), which asks how probable different causal structures are in light of the data. Implementing this approach requires specifying a prior. A highly tractable, and therefore common, choice is the Dirichlet-Multinomial specification, which assumes that each variable has finitely many possible realizations and places a Dirichlet prior over the conditional probability parameters \citep[see Section 17 of ][]{KollerFriedman}. Practically, inference then boils down to supplementing the observed data with hypothetical prior data (called \emph{pseudocounts}), and fitting the DAG to this combined dataset.

\subsubsection{Parameter learning with misspecified models}\label{sec:misspecifiedLearning}

As we have just seen, parameter learning is conditional on a causal structure. If that structure is misspecified, agents will be misled by the data even when learning is otherwise well-behaved. Therein lies a major appeal of this approach for behavioral economics. As \cite{Spiegler.QJE} writes,  the framework of fitting misspecified causal DAGs to data `\textit{provides a ``general recipe'' for transforming a standard rational expectations model into an equilibrium model with nonrational expectations.}'

Because many distinct DAGs can be fit to the same DGP, an immediate question is which misspecifications matter. Some subjective DAGs may serve as good enough `as-if' models, while others generate substantial errors.  The answer depends on whether the decision maker (henceforth \emph{DM}) is interested in using observations to form non-interventional beliefs, or whether she instead seeks to choose an intervention. A DAG's correlational implications suffice for the former case but not for the latter. We will discuss these two cases in turn.

\paragraph{Non-interventional beliefs with misspecified DAGs}

For the purpose of conditional predictions---that is, forming expectations conditional on observed variables---a causal DAG is completely characterized by the list of conditional independence relationships it implies (see Section \ref{sec:dagsIntro}). DAGs that represent the same list of conditional independence relationships are called \emph{Markov equivalent}. Therefore, as long as the agent's misspecified DAG is Markov equivalent to the DGP, her predictions will be just as good as they would be if she had fit the correct DAG.

If the agent fits a DAG that is not Markov equivalent to the DGP, she will make mistaken predictions for some nodes and some parameterizations of the DGP. However, these errors may be concentrated on variables of no direct interest to the agent. Is there a way to determine whether the agent's predictions will be wrong in a `relevant' way? In fact, there are two such results that differ by the definition of `relevant'.

First, \citet{Spiegler2020} considers a setting with $n$ nodes, $X_1, \ldots, X_n$. He assumes that the DM cares about the subset of variables $S\subseteq \{X_1, \ldots, X_n\}$. His Proposition 2 shows that the agent will have correct beliefs about the joint distribution of all variables in $S$ if and only if $S$ is an ancestral clique in some DAG $G$ that is Markov equivalent to the DM's subjective DAG. A set of nodes $S$ is an \emph{ancestral clique} in a DAG $G$ if $G$ directly connects all variables in $S$ to each other, and any variable that $G$ claims causes any other variable in $S$ is itself in $S$. To illustrate, in none of the DAGs in Figure \ref{fig:archetypalDAGs} do all three nodes form an ancestral clique, as each DAG involves two variables that are not directly connected. The sets $\{C,M\}$ in the chain (Panel A) and $\{C, E_1\}$ in the fork (Panel C) constitute ancestral cliques, but $\{C_1, E\}$ in the collider (Panel B) does not, since it includes $E$ while omitting one of $E$'s causes, $C_2$.

Second, as a corollary of the previous condition, \citet{Spiegler2020} considers a DM who is not interested in correlations between variables but merely cares about each variable's marginal distribution---for instance, if she is only interested in the mean of each variable. With this narrower definition of relevance, \citet{Spiegler2020} shows that $P_{subjective}(X_i) = P_{objective}(X_i)$ for all variables $i$ if the DM's subjective DAG is \emph{perfect}, that is, does not involve any $v$-colliders. Intuitively, someone viewing the data through a perfect DAG does not neglect the correlation between two variables that jointly cause a third. Such an agent cannot be `systematically fooled'---for instance made to think that the mean of any variable is higher than it actually is. \citet{Spiegler2020} additionally shows that if the DGP features a multivariate normal joint distribution of $X_1, \ldots, X_n$, then any subjective DAG the DM could fit to the data will imply the correct marginal distribution for each $i$, no matter whether it involves $v$-colliders or not.

The above results characterize \emph{when} DMs with misspecified subjective DAGs will make incorrect predictions, but not \emph{how large} their errors will be. \citet{Eliazetal.CheatingWithModels} answer the latter question in a multivariate normal setting by studying the extent to which fitting a misspecified model to data can distort estimates of pairwise correlations. The question is germane, for instance, to a DM seeking to interpret medical research about the effect of a particular behaviors, such as the consumption of table salt ($X_1$) on longevity ($X_3$). Because longevity is typically hard to observe, medical studies often rely on surrogate endpoints that are known to correlate with longevity, such as blood pressure ($X_2$). \citet{Eliazetal.CheatingWithModels} consider a DGP in which $X_1$ and $X_3$ are uncorrelated ($\rho_{1,3} = 0$) and ask what is the largest correlation that an analyst could infer by fitting the model $X_1\rightarrow X_2 \rightarrow X_3$? They show that for suitable joint distributions of $X_1, X_2$, and $X_3$, fitting this chain can yield an inferred correlation as high as 0.5 despite the absence of any true correlation between $X_1$ and $X_3$. In practice, such spurious inference can occur if researchers opportunistically select one from a range of plausible surrogate endpoints (\emph{marker hacking}). If the researcher can select multiple intermediate markers to create arbitrarily long causal chains to be fit, the inferred correlation between nodes that are in fact uncorrelated can be arbitrarily close to 1.

\citet{Spiegler.BehavioralCausalInference} considers the costs of a different type of error, namely the inclusion of the wrong set of control variables when agents seek to learn the causal effect of their actions on payoff-relevant outcomes from observational data. He finds that these welfare costs from estimating causal effects with bad controls differ greatly depending on whether the data from which agents learn are given exogenously or are endogenously produced by the agent's own actions.

\paragraph{Choice with misspecified DAGs}

When a DM seeks to predict the effects of her own choices or other interventions with the causal system, Markov equivalence to the DGP is no longer sufficient to rule out errors. Instead, the DM's subjective DAG must be \emph{identical} to that of the DGP. Otherwise, she will mispredict the effects of intervening on at least some variables for some parameterizations of the DGP. As in the case of non-interventional prediction, a DM with a misspecified DAG may mispredict only the effects of interventions that do not matter to her, perhaps because she does not have an opportunity to affect the relevant variables. For instance, the central bank can control interest rates but not government spending, and a manager may be able to influence the behavior of her subordinates but not that of her supervisors. Will some misspecifications be without consequence once we limit the set of nodes the agent can intervene on or cares about?

\citet{Spiegler.QJE} provides a precise answer to this question. Suppose the DM can intervene on a single node $A$ (his `action') and has a utility function defined over some subset $M$ of nodes. Even if the DM's DAG $G$ is misspecified, she will correctly perceive the mapping from her action to all utility-relevant outcomes if and only if the set of action and utility-relevant nodes, $\{A\}\cup M$, forms an ancestral clique in some DAG that is Markov equivalent to $G$, and the DM does not observe the realizations of any variables before intervening.

\paragraph{Equilibrium effects}

The above results apply regardless of how the data used for learning is obtained. Particularly interesting effects arise when the DM herself generates the data as a byproduct of her actions. In such settings, the use of a misspecified model can induce beliefs that justify taking those very actions, even if they are objectively suboptimal.

\cite{Spiegler.QJE}'s \emph{Dieter's Dilemma} illustrates the key idea. It considers the case of a person who mistakes a (harmless) symptom of a disease for its cause. The person has access to a dietary supplement that masks the symptom but does not affect the disease, and she only cares about the disease. That is, the true causal structure is \textit{Supplement} $\rightarrow$ \textit{Symptom} $\leftarrow$ \textit{Disease}, but the person believes \textit{Supplement} $\rightarrow$ \textit{Symptom} $\rightarrow$ \textit{Disease}. When she does not consume the supplement, she observes a strong correlation between the disease and the symptom, which she mistakes for a causal effect of the symptom on the disease. She thus infers that the supplement must be effective in treating the disease, and increases her supplement intake. If she always consumes the supplement, the symptom is masked; she observes no correlation between the symptom and the disease. She infers the supplement must be ineffective, and decreases her consumption. If her intake is high, she will lower it, and if it is low, she will raise it. Therefore, there will be an equilibrium in which she consumes the supplement with some intermediate regularity. The equilibrium consumption depends both on the magnitudes of the actual causal effects and on the costs and benefits associated with each action and outcome, and thus allows for rich comparative statics. If the symptom is a less reliable indicator of the disease, for example, the person believes the supplement to be less effective and thus consumes less of it in equilibrium. Like all models we discuss in this subsection, the Dieter's Dilemma assumes that the decision maker follows her subjective causal model dogmatically; she does not process the information about how medication intake and disease covary to falsify her misspecified DAG.\footnote{The existence of this interior equilibrium requires additional assumptions. The cost of the supplement must be sufficiently low, for instance, for the decision maker to ever want to consume it.}

\begin{figure}
\caption{Personal equilibrium}
\label{fig:personalEquilibrium}
\begin{center}
\includegraphics[width=0.7\textwidth]{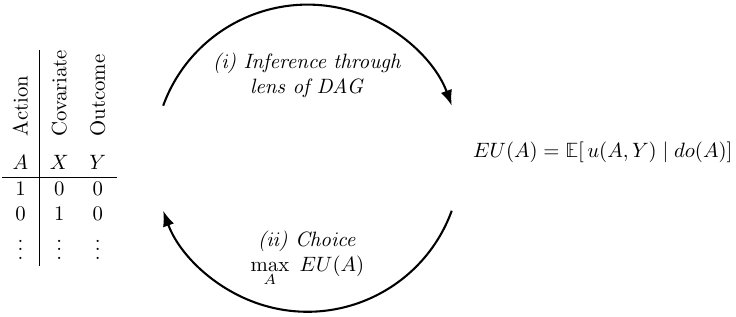}
\end{center}
\footnotesize
\textbf{\textit{Notes:}} Personal equilibrium requires that two conditions hold simultaneously. Condition 1: The joint distribution of all variables (`data') projected through subjective DAG justify the choices in condition 2. Condition 2: Choices produce the joint distribution of variables in condition 1.
\end{figure}
\iffalse % Original TikZ code for fig:personalEquilibrium
\begin{figure}
\caption{Personal equilibrium}
\label{fig:personalEquilibrium}
\begin{center}
\begin{tikzpicture}[>=Latex, line width=1pt]
    % Left: small table with extra top row (vertical labels)
    \node (table) at (0,2.2) {%
        $\begin{array}{c|cc}
        \rotatebox{90}{\text{Action}} & \rotatebox{90}{\text{Covariate}} & \rotatebox{90}{\text{Outcome}}\\[4pt]
        A & X & Y\\ \hline
        1 & 0 & 0\\
        0 & 1 & 0\\
        \vdots & \vdots & \vdots
        \end{array}$};

    % Top: inference text
    \node[align=center] (top) at (4.2,3.6) {%
        \textit{(i) Inference through}\\
        \textit{lens of DAG}};

    % Right: expected utility
    \node[align=left] (eu) at (9.2,2.3) {${EU}(A)=\mathbb{E}\!\left[\,u(A,Y)\mid {do}(A)\right]$};

    % Bottom: choice text
    \node[align=center] (choice) at (4.2,0.5) {\textit{(ii) Choice}\\ $\max\limits_{A}\ {EU}(A)$};

    % Big circular arrows
    \draw[->] (4.2,2.2) ++(160:2.6) arc (160:20:2.6);
    \draw[->] (4.2,2.2) ++(-20:2.6) arc (-20:-160:2.6);
\end{tikzpicture}
\end{center}
\footnotesize
\textbf{\textit{Notes:}} Personal equilibrium requires that two conditions hold simultaneously. Condition 1: The joint distribution of all variables (`data') projected through subjective DAG justify the choices in condition 2. Condition 2: Choices produce the joint distribution of variables in condition 1.
\end{figure}
\fi

The Dieter's Dilemma illustrates the concept of \emph{personal equilibrium} \citep{Spiegler.QJE}.\footnote{The concept has a similar flavor but is formally unrelated to  personal equilibrium in the literature on reference-dependent preferences \citep{koszegi2006model}.} It concerns an agent who interprets data through a given causal model. Her actions determine the data she observes, which she in turn uses to update the parameters of her subjective causal model. Personal equilibrium characterizes a situation in which the distribution of actions taken by the agent generates precisely the data which, when viewed through the lens of her model, justify taking that action distribution (see Figure \ref{fig:personalEquilibrium} for illustration). Formally, a probability distribution over actions is a personal equilibrium if any action that the DM takes with positive probability maximizes her subjective expected utility according to the beliefs generated from fitting her subjective DAG to the data.

Personal equilibrium yields interesting applications that we discuss in Section \ref{sec:applications}. Besides generating novel insights, personal equilibrium also subsumes well-known behavioral frameworks as special cases. For example, \cite{eyster2005cursed}'s \emph{cursed equilibrium} considers a specific form of misspecification in which players in a game neglect the information content of others' actions. In a market for lemons \citep{akerlof1970market}, a `cursed' buyer ignores the fact that a seller has private information about the value of a car. This buyer can be described as processing the decision problem through a DAG that omits the link from the seller's knowledge about the car to the seller's willingness to accept. Personal equilibrium is also closely related to \emph{$S(K)$-equilibrium} \citep{OsborneRubinstein1998ProcedurallyRational}, \emph{analogy-based equilibrium} \citep{jehiel2005analogy}, and \emph{naive behavioral equilibrium} \citep{Esponda2008BehavioralEquilibrium}. It is a special case of \cite{esponda2016berk}'s \emph{Berk-Nash equilibrium}, which does not restrict misspecification to (unparametrized) DAGs but allows for general misspecified likelihood functions.

Even with misspecified DAGs, personal equilibrium effects do not always arise. When changing the frequency of the action leaves the agent's perceived mapping from actions to consequences unchanged---a property \citet{Spiegler.QJE} calls \emph{consequentialist rationality}---there is no feedback loop and the agent's optimization problem is standard. \citet{Spiegler.QJE} shows that this property holds whenever every variable that the agent does not treat as a direct consequence of her action is, under the true DAG, independent of the action given the causes the agent does recognize. When consequentialist rationality is violated, \citet{CarbajalNachbar} show that the resulting personal equilibrium effects are generic; they arise for all but a measure-zero set of distributions consistent with the true DAG.

While personal equilibrium considers agents whose models are structurally misspecified but correctly fit to the data, a complementary literature considers agents whose models are structurally correct but impose erroneous parametric restrictions. This approach has been used to explain phenomena such as misdirected learning, stereotyping, and bias substitution \citep{Heidhuesetal.2018, Heidhuesetal.2025}.

\subsection{Empirical evidence}\label{sec:parameterLearningEmpirics}

We next turn to the empirical question of how humans interpret data when they have beliefs about the structure of causal relationships between observable variables but not their strengths. To answer this question, we largely retain the empirical literature's focus on binary variables.

The literature has considered various definitions of the magnitude of a causal effect. The quantity known as $\Delta P$ \citep{JenkinsWard1965} appears naturally in economic decision making. Consider an agent who decides whether to spend amount $k$ to bring about a cause $C$ (such as a treatment for an illness, or an investment in education) to influence an outcome $E$. Doing so is optimal if 
\bmath \label{deltaPutility}
\sum_{e\in\{0,1\}} \big(P(E=e|\d(C=1)) - P(E=e|\d(C=0))\big) U(E=e) \geq k
\emath 
where $U$ denotes the agent's money-metric utility over $E$, which we assume is separable from the expense $k$.
The measure of causal strength is simply defined as the change in the chance the effect obtains if the cause $C$ is exogenously imposed present rather than absent, $\Delta P = P(E=1|\d(C=1)) - P(E=1|\d(C=0))$, which appears as the first factor in equation (\ref{deltaPutility}).\footnote{While equation (\ref{deltaPutility}) refers to interventional probabilities, the original formulation of $\Delta P$ by \citet{JenkinsWard1965} uses observational probabilities. The two cases coincide if $C$ is exogenous.} 

The key economic question regarding parameter learning is thus what individuals infer about $\Delta P$ in a given setting based on observable data and their beliefs about the causal structure of the DGP. We review the empirical research on this question in four steps. First, we consider learning when the data features causal effects, focusing on whether and how subjective probabilities are warped versions of objective probabilities. Second, we consider the important special case in which the DGP shows no causal effect, where subjects may nonetheless perceive non-zero causal effects. Both cases suggest that the apparently biased mapping from objective probabilities to subjective judgments arises in part because subjects reason as if unobserved background causes and hidden interactions are also at work. Third, we discuss research on the priors that people tend to use in parameter learning, which, if they are strong and data is finite, can bias their posteriors. Fourth, we discuss research that focuses on cases in which individuals' mental models may be structurally misspecified, both when they simply respond to observational data, and when they learn from the data they generate themselves, as in \citeapos{Spiegler.QJE} personal equilibrium.

\paragraph{Learning non-zero associations}\label{sec:elementalCausalInduction}

How well do subjects learn the magnitudes of existing causal associations? \citet{AmbuehlHuang}'s data provides an answer in a setting with binary variables. They regress subjects' reported beliefs about causal effects $P(Y|\d(X=1,Z))-P(Y|\d(X=0,Z))$, where $Z$ may be empty, on the causal effects observed in the data. They find substantial attenuation, with many subjects perceiving effects merely a quarter as large as the true association. By contrast, in a setting with continuous predictor and outcome variables---which entail much more information than binary realizations---\citet{Rehder2025} finds that subjects' average estimates of the slope parameters linking variables are statistically indistinguishable from their actual values, though there is substantial heterogeneity across individuals.

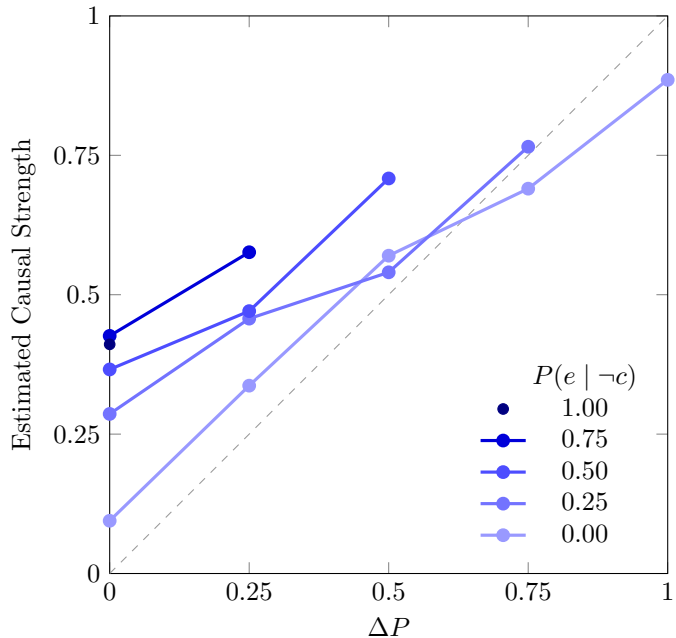
\begin{figure}[htb]
\caption{Empirical mapping of objective probabilities into Likert-scale ratings of `causal strength'}
\label{fig:buehneretal}
\begin{center}
\definecolor{navy}{RGB}{0,0,128}
\begin{tikzpicture}
\begin{axis}[
    width=9cm,
    height=9cm,
    xlabel={$\Delta P$},
    ylabel={Estimated Causal Strength},
    xmin=0, xmax=1,
    ymin=0, ymax=1,
    xtick={0,0.25,0.5,0.75,1},
    ytick={0,0.25,0.5,0.75,1},
    legend pos=south east,
    legend style={draw=none},
    legend reversed=true,
]
% Diagonal
\addplot[black!40, thin, dashed, forget plot, domain=0:1, samples=2] {x};
% P(e|¬c) = 0.00 — lightest
\addplot[blue!40!white, mark=*, solid, very thick, mark size=2pt] coordinates {
    (0.00,0.0946)
    (0.25,0.3369)
    (0.50,0.5700)
    (0.75,0.6902)
    (1.00,0.8854)
};
\addlegendentry{$0.00$}
% P(e|¬c) = 0.25
\addplot[blue!55!white, mark=*, solid, very thick, mark size=2pt] coordinates {
    (0.00,0.2862)
    (0.25,0.4571)
    (0.50,0.5402)
    (0.75,0.7654)
};
\addlegendentry{$0.25$}
% P(e|¬c) = 0.50
\addplot[blue!70!white, mark=*, solid, very thick, mark size=2pt] coordinates {
    (0.00,0.3660)
    (0.25,0.4706)
    (0.50,0.7085)
};
\addlegendentry{$0.50$}
% P(e|¬c) = 0.75
\addplot[blue!85!black, mark=*, solid, very thick, mark size=2pt] coordinates {
    (0.00,0.4262)
    (0.25,0.5762)
};
\addlegendentry{$0.75$}
% P(e|¬c) = 1.00 — darkest (navy)
\addplot[navy, only marks, mark=*, mark size=2pt] coordinates {
    (0.00,0.4110)
};
\addlegendentry{$1.00$}
\addlegendimage{empty legend}
\addlegendentry{$P(e\mid \neg c)$}
\end{axis}
\end{tikzpicture}
\end{center}
\footnotesize
\textit{\textbf{Notes}}:
Results from \citet{BuehnerChengClifford2003}. Treatments with constant $P(e\mid \neg c)$ in the data are connected by lines. The horizontal axis displays $\Delta P = P(e\mid c) - P(e\mid \neg c)$.
\end{figure}

While few papers put objective causal effects and their subjective counterparts on a directly comparable scale, a large literature in psychology on \emph{elemental causal induction} studies how objective causal effects map onto Likert-scale ratings of `causal strength' that are often not precisely defined to subjects.\footnote{For instance, \citet{BuehnerChengClifford2003} ask subjects to `\textit{judge how strongly each vaccine prevented the disease related to the virus in question by giving a rating on a scale from 0 (the vaccine does not prevent the disease at all) to 100 (the vaccine prevents the disease every time).}'} Hence, nonlinear or otherwise surprising mappings between such ratings and objective causal effects can arise either from biased inference or from interpreting the questions as eliciting something other than probabilities. Leaving this interpretational ambiguity unresolved, the literature finds that subjects' Likert-scale statements about causal strength strongly covary with $\Delta P$, but systematically deviate from it depending on the probability with which the effect is present even when the cause is absent. Figure \ref{fig:buehneretal} shows a typical result. \citet{BuehnerChengClifford2003} argue that these judgments are well-described by the measure $q=\Delta P / (1-P(E|\neg C))$, termed \emph{Power PC} by \citet{cheng1997covariation}.

The fact that \emph{Power PC} is not the relevant measure for economic decision making leads one to ask why it would describe subjects' statements. The seminal contribution of \citet{griffiths2005structure} argues that the reason is humans' tendency for causal rather than purely statistical thinking. They assume that individuals do not treat the effect's occurrence in the absence of the causes as just a statistical statement, but instead posit an alternative unobserved background cause $B$ that could explain the effect. They show that if subjects assume the effect $E$ is a common consequence of the cause $C$ and the background cause $B$, linked through a noisy-OR specification, then the implied causal strength parameter for $C$ is given by the Power PC measure. In a similar way, $\Delta P$ also corresponds to the implied strength parameter under a linear specification. It remains an open question whether humans think about causal strength in terms of Power PC even in economic decision making, when doing so can reduce utility, or whether Power PC merely describes their responses to survey questions.

\paragraph{Failure to learn the absence of associations}

While individuals sometimes perceive actual effects in attenuated form, they also show the opposite bias---perceiving correlations and causal effects where none exist \citep[reviewed in][]{FiedlerReview, Matuteetal.Review, fiedler2022illusory}. In a typical experiment on such \emph{illusory correlation}, subjects observe, for instance, data on 120 patients who differ in whether they received a medication and whether they recovered, with the frequencies shown in Panel A of Table \ref{tab:illusoryCausation}. From this data, the average subject typically infers that the medication is effective, even though the probability of recovery is 80\% regardless of medication intake. The effect appears when both the effect and the cause have skewed marginal distributions. Subjects behave as if common causes are associated with common outcomes, and rare causes with rare outcomes. When interpreted causally, illusory correlation has interesting implications for research on stereotyping, as one can see by relabeling the contingency table as Panel B of Table \ref{tab:illusoryCausation}. Subjects typically infer from such data that minority members commit crimes more often than majority members, even though it shows no such association \citep{HamiltonGifford1976}.

\begin{table}[htb]
\caption{Illusory causation: Example contingency tables}
\label{tab:illusoryCausation}
\begin{center}
\textbf{A. Medical example} \\
\begin{tabular}{lcc}
\toprule
                     & Recovered & Did not recover \\
\midrule
Took medication      & 80        & 20              \\
Did not take medication & 16     & 4               \\
\bottomrule
\end{tabular}
\\ \bigskip
\textbf{B. Stereotypes} \\
\begin{tabular}{lcc}
\toprule
                     & Did not commit crime & Committed crime \\
\midrule
Majority member      & 80        & 20              \\
Minority member & 16     & 4               \\
\bottomrule
\end{tabular}
\end{center}
\end{table}

While illusory correlation may seem like a mistake, recent research argues that it is not necessarily irrational. One strand of research \citep{costello2019rationality, bott2021normative, GershmanCikara} shows how it can arise from Bayesian updating. \citet{GershmanCikara}, for instance, consider two groups with equal true means. If the assessor’s Gaussian prior has a mean below the sample average and one group is observed more often, posterior means (precision-weighted averages of prior and data) will differ: the larger group’s estimate is pulled closer to its empirical mean, yielding a higher posterior. A second strand attributes it to causal elaboration mechanisms. Following a similar argument to \citeauthor{griffiths2005structure}'s (\citeyear{griffiths2005structure}) rationalization of the Power PC measure of causal strength, \citet{NgLeeLovibond2026Cognition} argue that people who observe the effect occurring in absence of the cause may posit background causes that vary across conditions and thereby account for the effect instead. Indeed, in a medical scenario similar to that of Panel A of Table \ref{tab:illusoryCausation}, their participants spontaneously assumed that patients who were not given the medication had stronger immunity or less severe illness than those who were given it. Positing this additional mechanism let them reconcile their belief in a positive causal effect with the statistical independence observed in the data. 

The type of illusory causation just discussed arises purely from interpreting contingency tables, regardless of whether the subject merely observes data or takes actions in an attempt to influence outcomes. The \emph{illusion of control} is a variant of this phenomenon that is likely not driven by illusory correlation. It occurs in situations such as die tossing, where subjects know that affecting outcomes is impossible, but still behave as if they could affect them \citep{langer1975illusion, langer1975heads}. In cases where mere observation yields illusory causation, personal involvement does not strengthen perceived contingencies \citep{ion2014illusion}, though it does distort which variables subjects include in their causal models \citep{fan2024choice}.

\paragraph{Priors}

Even if causal effect magnitudes can be learned correctly when given enough data, priors about parameters will affect beliefs when samples are finite. What priors do individuals tend to bring to new problems? 

\citet{Lu2008GenericPriors} argue that human priors do not follow the uniform distribution across parameters that some papers assume for technical convenience \citep[e.g.,][]{griffiths2005structure}. Instead, they argue that people `prefer causal models that minimize the number of causes ... while maximizing the strength of each individual cause that is in fact potent.' Empirically, they showed that a formalization of such \emph{strong and sparse} priors, when updated with data, better fits human posterior judgments than does updating uniform priors. To measure priors more directly, \citet{Yeung2015Expectations} used an experimental design based on iterated learning. Under the assumptions of their model, this process defines a Gibbs sampler in which subjects' choice distributions would converge to their priors. Using this method, they confirmed that subjects' priors favored near-deterministic relationships between the main cause and the outcome. In contrast to \citet{Lu2008GenericPriors}'s proposal, however, their empirically identified priors showed no preference for the probability of the main cause $C$ over alternative causes, and no competition between the probabilities of the main and background causes. The preference for {(near-)determinism} in people's priors appears to vary across individuals but stable for a given individual \citep{Mayrhofer2015SuffNec}.

\paragraph{Learning with misspecified causal models}

Two different strands of empirical research focus on learning and choosing with misspecified models. 

The first strand tests the main qualitative assumption of the misspecified models literature: do people with different subjective models interpret the same data differently?  
Because these studies seek to make a broad, qualitative point, they are not always set in the context of causal DAGs. \citet{BarronFries} show that externally provided interpretations for a series of ten data points (specifically, claims about a structural break in the DGP) significantly affected subjects' predictions, even when subjects knew that the information providers had incentives to bias their interpretation against the subjects' own interest. \citet{CharlesKendall} specifically focus on DAGs. They consider the setting of \citet{EliazSpiegler.Narratives} in which subjects observe a table listing realizations of triples of binary variables $A, X, Y$. Their results show that subjects' choices could be biased by providing them with interpretations that suggested that the data were consistent with a chain $A\to X \to Y$ or a collider $A\to X \gets Y$.\footnote{These interpretations take the form `\textit{$X=1$ only if $A = 1$. Further, when $X=1$, $Y=1$ is always true. So, choose $A=1$ more often than $A=0$}' and `\textit{If $X=0$, it always holds that $Y=0$ if $A=1$. To counteract this, choose $A=0$ more often than $A=1$ so that $Y$ can be 1 even if $X=0$.}'} A caveat is that all interpretations included directives like `\textit{So, choose $A=1$ more often than $A=0$}' which produced similar behavioral effects even when presented without any explanation of the data.

The second strand of research tests the specific predictions of the theory that individuals interpret data as if fitting a mixture of causal DAGs to it. Specifically, \citet{AmbuehlHuang} tested whether subjects' causal belief systems were rationalizable as arising from interpreting data through the lens of a causal DAG or a mixture of DAGs. They also tested the predictions of \citet{Spiegler.QJE}'s personal equilibrium concept, conducting three types of tests. First, for the Dieter's Dilemma, they derive a theoretical result showing that the only subjective DAG that can rationalize believing in the ability to affect the outcome is the chain, $A\to X \to Y$. Any subject who interprets the data through this DAG, however, must also believe that $X$ blocks the path from $A$ to $Y$. \citet{AmbuehlHuang} found that this prediction was empirically violated. Second, acknowledging that model uncertainty may cause subjects to behave as if by interpreting the data through a mixture of DAGs, they identified a set of restrictions that beliefs consistent with this theory must satisfy. Their data also rejected these restrictions. Third, they tested the comparative statics predictions of Personal Equilibrium, for instance by varying the parameters of the DGP to increase the equilibrium action frequency, or by altering the link function to produce multiple equilibria. The data directionally matched the predicted comparative statics, but these effects were substantially attenuated. This attenuation may reflect factors such as choice stochasticity, which could more generally weaken the model's predictions across settings in which \citet{Spiegler.QJE}’s model applies.

\section{Structure learning} \label{sec:structureLearning}

In the previous sections, we treated the structure of the causal model as given, focusing on how agents reason within that structure and estimate its parameters. However, this presupposes that agents hold beliefs about which structure describes the causal relationships between variables in the first place. How should idealized agents acquire causal structure, and how do humans actually do so?

A natural question is why agents would seek to infer structure rather than simply estimating parameters in a single, fully connected model. The answer is that structure constrains inference, and thereby makes learning more efficient. By ruling out many possible relationships among variables, learners can draw stronger conclusions from limited data than a completely unstructured approach would permit.

In this section, we first review normative accounts of structure learning (Section \ref{sec:structureLearningConcepts}), including constraint-based and Bayesian frameworks, as well as hierarchical formulations that allow structure to be shared across contexts. We then examine empirical evidence (Section \ref{sec:structureLearningEvidence}) on how accurately individuals infer causal structure, the conditions under which they succeed or fail, and the processes by which they update beliefs about structure.

\subsection{Theoretical concepts}\label{sec:structureLearningConcepts}

\subsubsection{Constraint-based learning}

A frequentist approach to inference about causal structure is \emph{constraint-based} learning \citep{spirtes2000causation}. The main idea is to use the set of conditional independence relations implied by each causal DAG to narrow down the set of candidate structures, ideally to the Markov equivalence class of the DGP's causal structure. For example, the chain and fork structures in Panels A and C of Figure \ref{fig:archetypalDAGs}, relabeled to $X \to Y \to Z$ and $X \gets Y \to Z$, respectively, imply that $X$ and $Z$ are independent conditional on $Y$. By contrast, the $v$-collider of Panel B, relabeled to $X \to Y \gets Z$, implies that $X$ and $Z$ are unconditionally independent. Hence, if a dataset of observations of $X, Y$, and $Z$ reveals independence between $X$ and $Z$ conditional on $Y$ but no other independence relations, this observation is consistent with the chain $X\to Y \to Z$ and the fork $X\gets Y \to Z$ (as well as the reverse chain $X\gets Y \gets Z$), which form a Markov equivalence class, but not with other DAGs. If, on the other hand, the dataset shows independence of $X$ and $Z$, but dependence between all other pairs of variables, both  conditionally and unconditionally, the only compatible three-node DAG is the $v$-collider of Panel B. In practice, independence is asserted based on null hypothesis significance tests. In settings with more than a handful of variables, the vast number of possible DAGs requires specialized algorithms to discover the Markov equivalence class of the DAG consistent with the list of conditional independencies.

\begin{figure}[htb]
\caption{Bayesian structure learning example.}
\label{fig:structLearning}
\centering
\includegraphics[width=0.85\textwidth]{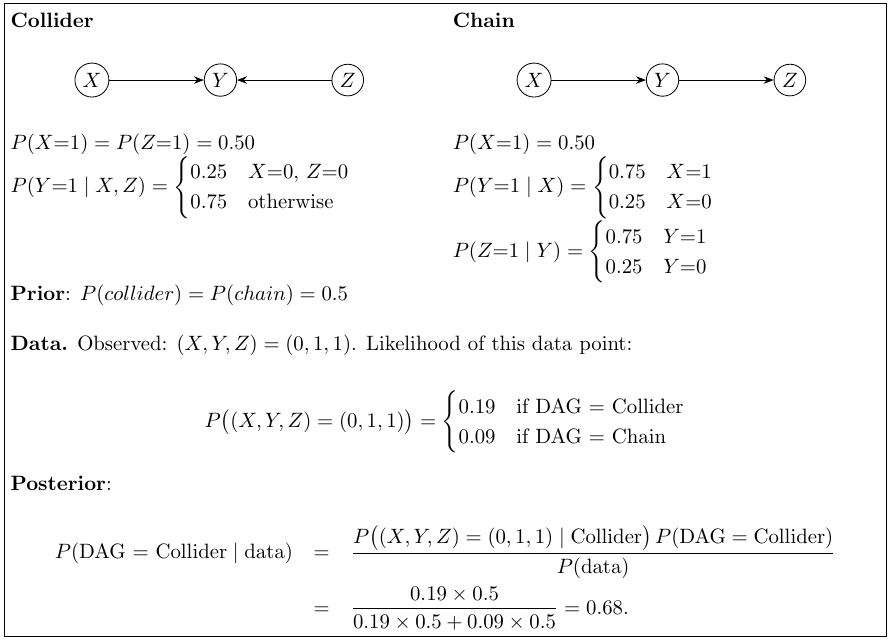}
\footnotesize
\flushleft
\textbf{\textit{Notes}}: In this example, the learner has no uncertainty about the parameters of the DAG. In general, the learner faces a distribution over parametrizations of the DAGs. She uses data to update not only about the likelihoods of the structures, but also about the parameters within each structure.
\end{figure}

\subsubsection{Bayesian structure learning}

The cognitive science literature more commonly models structure learning in a Bayesian framework. This approach offers several advantages, some of which make it particularly appealing as a model of human judgment. First, Bayesian methods quantify uncertainty by assigning posterior probabilities to alternative causal structures. Constraint-based methods, in contrast, simply output a set of DAGs deemed consistent with the data. Second, the Bayesian framework extends readily to hierarchical settings in which different environments share regularities (the same physical laws, for instance, govern both a bike rolling downhill and a door swinging shut). This makes it well-suited to capture how people learn abstract higher-level patterns and generalize across contexts. Third, it can handle data of different types---combining, for example, historical observations with verbal explanations. Fourth, it naturally accommodates individual differences in prior beliefs, allowing agents to assign different posteriors to DAGs that are observationally similar. Fifth, Bayesian models have inspired approximate process accounts meant to better incorporate the cognitive constraints of human learners.

In principle, the mechanics of Bayesian structure learning are straightforward: calculate the probability of the observed data for each structure, then use them in Bayes' law to update beliefs about structures. Figure \ref{fig:structLearning} shows an example. A learner entertains two candidate models: a collider DAG and a chain DAG. For simplicity, suppose the learner knows the parameters of each structure, and assigns equal prior probability to the two DAGs. How should she update her beliefs upon observing the data point $(X, Y, Z) = (0,1,1)$? To answer this question, recall that a parametrized DAG implies a probability distribution over all possible outcomes. Hence, we can calculate each candidate DAG's probability of producing observation $(X, Y, Z) = (0,1,1)$. These probabilities, along with the prior probability of each model, determine the posterior probability of each model according to Bayes' law. In this specific example, the observation increases the learner's belief that the DGP is the collider DAG from 0.5 to 0.68.

The above example illustrates the general principle---a simple restatement of Bayes' law---that the posterior probability of a model given the data is proportional to the probability of the data given the model,
\bmath \label{eqn:bayesStructure}
P(\textrm{model} | \textrm{data}) \propto P(\textrm{data} | \textrm{model})P(\textrm{model}).
\emath
If the parameters of each structure are also uncertain, we calculate the likelihood that a given structure produced the data by averaging over the distribution of possible parameter values: 
\[P( \textrm{data} | \textrm{model}) = \sum_{parameters}P(\textrm{data} | \textrm{model, parameters})P(\textrm{parameters} | \textrm{model}).\]

The two factors on the right-hand side of equation (\ref{eqn:bayesStructure}) highlight important properties of structure learning. The likelihood, $P(\textrm{data} | \textrm{model})$, captures an intuitive feature of how people learn structure; upon observing data, we increase our weight on the structures that are more likely to have produced it and decrease our weight on those that are less likely to have produced it. More surprisingly, it embeds a preference for simplicity in structure learning via the so-called \emph{Bayesian Occam's Razor}. Consider a simple model and a more complex one, and suppose both can explain a given set of observations similarly well. More complex models typically admit a wider range of parameter values, many of which fit the observed data poorly, but the marginal likelihood averages over all parameter values consistent with a model. As a result, the complex model is penalized more heavily, and beliefs update more strongly toward the simpler model \citep{mackay2003information}.\footnote{While the Bayesian Occam's Razor has rarely been the explicit focus of work on structure learning, human cognition shows consistency with some of the hallmarks of Bayesian Occam's Razor in other settings \citep{blanchard2018bayesian}.}

Priors impose ex ante tendencies on structure learning, and can be formulated either directly over the DAGs themselves via $P(model)$, or indirectly over the parameters \citep{Lu2008GenericPriors} or functional form \citep{Lucas2010HierBayes} conditional on a DAG. By encoding assumptions about the typical strength, sparsity, or determinism of causal relationships, prior beliefs influence which structures appear most plausible. For instance, beyond the tendency to favor simpler models induced by Bayesian Occam's Razor, a preference for simpler or more complex structures may stem from learners' priors over DAGs, as when those priors favor coarser partitions of variables \citep{Kemp2010LearnCausalModels}.

Once an agent has drawn inferences about the likelihood of multiple structures, how does she make predictions or choose between actions? There are two common approaches \citep{KollerFriedman}. \emph{Model selection} entails picking the single most probable structure (e.g., the \emph{maximum a posteriori}, or MAP, estimate) and choosing the action implied as optimal by that structure. In \emph{model averaging}, by contrast, learners weight predictions about the effects of interventions by each structure's posterior probability and choose the action that yields the best outcome according to these weighted predictions.

While the data in the example above comprises a vector of binary outcomes, the Bayesian framework can accommodate and integrate data of any kind. For instance, information is often communicated through verbal explanations \citep{sloman2009causal, lombrozo2017causal, meder2017inference}, which can be cast as probabilistic evidence within a Bayesian model \citep{nam2023show}, if one provides an explicit stochastic mapping from causal structures to natural language statements. While straightforward in principle, implementation is challenging in practice because this mapping must operate over the vast space of possible utterances \citep{barashersiegal2026}.

\subsubsection{Bayesian Hierarchies and the Blessing of Abstraction}

The Bayesian approach to structure learning raises the question of where priors about causal structure come from. In general, they may reflect past experience, background assumptions, or abstract knowledge carried over from other domains. \emph{Hierarchical Bayesian models} provide an influential formal account of how priors can be learned from experience across settings.

Hierarchical Bayesian models extend ordinary Bayesian inference by allowing priors themselves to depend on higher-level variables that are also learned from data \citep{griffiths2009theory}. In the case of structure learning, this means treating the causal structure of a particular system as a draw from a higher-level distribution over structures, which is itself inferred from experience. At the lower level, the learner infers the causal structure governing the system at hand---such as how the success of a specific firm depends on factors like location and product quality. At the higher level, the learner infers regularities about what kinds of causal graphs are typical across systems in a domain---for example, learning that firms generally succeed due to multiple interacting causes rather than a single dominant factor. These hierarchies can be extended arbitrarily---at a still higher level, one might infer that social and economic systems like firms tend to be described by densely connected graphs, whereas many physical systems are represented by more sparsely connected graphs.

\citet{Kemp2010TheoryFormation} illustrate these ideas in a setting involving biomedical data. Their hierarchical Bayesian model receives a set of entities (e.g., ``lung cancer,'' ``ammonia,'' ``fever'') and observed relational predicates between them (e.g., ``causes,'' ``affects''), with no labels indicating what kind of entity anything is. At the higher level, the model learns a framework theory specifying categories such as chemicals, diseases, symptoms, organisms, as well as the laws that connect them (e.g., chemicals cause diseases, diseases produce symptoms). At the lower level, these categories and laws constrain inferences about specific relationships between individual entities, such as whether asbestos causes lung cancer. Even though neither level is given in advance; their model successfully learns the categories, laws, and specific relational structure from raw data. While \citet{Kemp2010TheoryFormation} is about discovering relational systems of concepts in general, \citet{Kemp2010LearnCausalModels} study the same hierarchical logic in an explicitly DAG-based setting. At the lower level, their model learns the causal properties of individual objects, while at the higher level, it learns which kinds of objects tend to produce which kinds of effects. Because the model captures abstract regularities shared by different objects, it can infer the properties of a new object from as little as a single observation. More fundamentally, \citet{Goodman2011Theory} show that even the abstract principles of the causal DAG framework, such as the fact that dependence is directed and that interventions are exogenous, can be learned at a higher level, while specific causal graphs are learned at a lower level.

The hierarchical modeling approach thus addresses the provenance of priors by treating them as the product of Bayesian inference at higher levels. It also explains how learning transfers between contexts. Observations in one case shape beliefs about others insofar as they inform higher-level inferences shared across contexts. 
This feature of hierarchical models can give rise to what is known as the \emph{Blessing of Abstraction} \citep{Goodman2011Theory, gershman2017blessing}. In contrast to the \emph{Curse of Dimensionality} which makes unconstrained learning over high-dimensional distributions impractical, hierarchical inference allows learners to pool information across environments and extract abstract regularities supported by a larger effective sample size than any single instance could provide. Once learned, these abstractions narrow down the hypothesis space in new contexts and thereby accelerate inference. Hence, the blessing of abstraction makes hierarchical Bayesian inference exceptionally potent, and helps explain how people can learn so much from what seems like so little data.

\subsection{Empirical evidence}\label{sec:structureLearningEvidence}

The empirical literature on structure learning is vast. We first review evidence on the question of whether and how well subjects can learn structure at all (Section \ref{sec:dataLimitations}), and whether their beliefs display the typical hallmarks of constraint-based (Section \ref{sec:constraintBasedLearning}) or Bayesian learning (Section \ref{sec:empiricalBayesianLearning}). Because structure learning can be computationally intensive, we then consider some processes by which humans are thought to approach Bayesian solutions, as well as deviations that stem from cognitive limits (Section \ref{sec:processesOfStructureLearning}).

\subsubsection{How well do people learn structure?} \label{sec:selection-data}\label{sec:dataLimitations}

Broadly speaking, empirical research in cognitive science shows that people are able to learn causal structure from data, though perhaps more slowly and less accurately than ideal learners \citep{Rottman.HandbookChapter}.
For example, participants in \citet{Coenenetal.InterventionStrategies} were able to select the correct three-node DAG from two options 87\% of the time when allowed to freely intervene on the DAG \citep[see also][]{Coenenetal2019}. When given finite observations, various studies find that participant performance matches Bayesian benchmarks \citep[e.g.,][]{nam2023show}. Even children as young as four years old draw correct causal inferences from observing interventions in simple cases \citep{gopnik2001causal, GopnikEtAl2004}. Related evidence from economics points in the same direction. \citet{Frechetteetal.} presented subjects with datasets of triples of binary variables. Subjects studied these datasets knowing that they would later have to predict variables based on observations of other variables in the same system using only the notes they had taken, without access to the original dataset. Subjects' largely successful predictions indicated considerable efficacy in extracting the DAG structure of the DGP.

In addition to such conscious, deliberate learning, structure inference occurs at an automatic, perceptual level \citep[reviewed in][]{ShamsBeierholm}. The \emph{ventriloquist illusion}, a canonical example of \emph{sensory cue integration} \citep{kording2007causal, sato2007bayesian}, illustrates this point. Observers must infer where a voice comes from when they both see a puppet and hear `its' voice. The perceptual system appears to consider two possibilities: that the voice comes from the puppet’s \emph{seen} location (common cause), or from somewhere else (separate causes). When what is heard is near what is seen, the common cause model gets more weight, and the voice is perceived as coming from where the puppet is \emph{seen}. As the mismatch increases, the separate cause model gets more weight, and the perceived source shifts back toward where the voice is \emph{heard}. This non-monotonic pattern in the distance between the perceived and actual location of the voice as a function of the distance between the puppet’s seen location and the location of the voice is a signature of Bayesian model averaging.

At the same time, a large body of work documents substantial failures of causal structure learning. For instance, the subjects in \citet{LagnadoSloman2004} performed no better than chance when asked to pick the correct causal structure from five options after observing realizations from a three-variable probabilistic system for 50 trials. Similarly, the subjects in \citet{White2006CausalStructure} performed near chance levels when seeking to infer structure from observational data as written sentences describing patterns of co-occurrence between populations of species in a nature reserve. \citet{Fernbach2009Local} also document poor structure learning from observation. Indeed, performance can vary greatly within the same studies \citep[e.g.,][]{LagnadoSloman2002, Steyversetal.}.

What explains this variation in the success of causal structure learning? Performance is typically higher when subjects can actively intervene \citep{LagnadoSloman2002, Coenenetal.InterventionStrategies}, when information about temporal order or delay is available \citep{lagnado2006time, BramleyGerstenbergetal2018}, when strong domain priors or hierarchical structure guide inference \citep{griffiths2009theory, Kemp2010LearnCausalModels}, when the system is deterministic or low-noise \citep{MayrhoferWaldmann2016, Rothe2018Successful, Frechetteetal.}, and when each effect has few causes \citep{Rothe2018Successful}. In short, success is more likely when the available information is richer and the underlying structure is simpler or less noisy.

\subsubsection{Constraint-based learning}\label{sec:constraintBasedLearning}

While cognitive science largely focuses on settings suited to Bayesian structure learning, some economic applications are better captured by constraint-based learning. Consider agents who choose among externally proposed models (for instance, because they arise in public policy debates), possibly by comparing them to statistical summaries of historical data. \citet{AmbuehlThysen} study this case in a stylized four-variable environment where subjects could inspect aggregate statistics of the DGP---that is, any conditional or unconditional correlation among pairs or triplets of variables. Subjects were shown pairs of candidate models, each recommending a different investment level, and chose which to follow. Since payoffs depended on the chosen investment and the true DGP, they faced an incentive to identify whether  the correctly specified model was in the choice set, and, if so, select it. The models were described in natural language (e.g., `$X$ directly affects $Y$,' `$X$ influences $Y$ indirectly through $Z$,' or `$X$ is a symptom of $Z$') and illustrated with DAGs. They find that individuals displayed a remarkable ability to discard misspecified models; three fifths of their subjects did so consistently. Success in this task required not only intuiting the relevant correlational implications but also finding which of 18 possible charts displaying empirical data would help check these implications. Subjects' performance stemmed from the possibility of relying on qualitative (i.e., whether model-predicted correlations coincide with those observed in the data) rather than quantitative inference (i.e., how strongly investments and payoffs covary in the data). In fact, subjects rarely even viewed the data necessary for quantitative inference. Moreover, failures to detect misspecified models reflected limited ability rather than limited motivation, as evidenced by the fact that tripling the monetary stakes (up to \$90 in some cases) had no effect.

\subsubsection{Hallmarks of Bayesian structure learning}\label{sec:empiricalBayesianLearning}

Other applications may be better described by Bayesian structure learning, especially those in which agents learn from the piecemeal arrival of individual observations. Beyond the fact that people routinely draw causal inferences from samples too small for standard statistical tests to detect reliable dependencies \citep{gopnik2001causal}, several lines of evidence indicate that human learning bears several of the distinctive features of Bayesian structure inference. Among these are findings that people simultaneously entertain and update weights on multiple causal structures, that their inferences reflect prior beliefs about the domain, and that these priors can themselves be updated with experience.

\paragraph{Simultaneous inference over competing structures}
Evidence from \citet{Gershman2017Context} suggests that learners do not commit to a single causal structure, but instead maintain and update beliefs over multiple candidates. In the experiment, participants learned relationships among three variables: the cue, the context, and the outcome. In one condition, context was irrelevant; in a another, it could affect the outcome probability independently; in a third, it modulated the effect of the cue on the outcome. Subjects then made predictions for combinations of cues and contexts, including in cases where the context variable took values not seen before. When the context was irrelevant, participants inferred that the cue signaled the outcome regardless of context; when the context had an independent effect, participants inferred that its effect on the outcome did not depend on the cue; but subjects were reluctant to generalize when cue and context both mattered together. This pattern is parsimoniously explained by a Bayesian model that assigns probabilities to the candidate structures based on the observations. Related evidence comes from sequential decision-making tasks studied by \citet{AcunaSchrater2010}, where participants' behavior was consistent with maintaining a posterior over causal structures rather than committing to a single one. In these tasks, participants had to learn whether rewards across options were independent or coupled. When the options were likely independent, subjects explored to gather information about each option, whereas when they were likely coupled, they exploited more aggressively because data on one option was informative about the other. When the evidence was ambiguous, behavior lay between these extremes, changing smoothly with the inferred likelihood of each structure. Though these studies do not directly measure beliefs over structures \citep[cf.][]{Steyversetal.}, they are consistent with the idea that learners simultaneously entertain multiple causal models and update their relative plausibility over time.\footnote{These papers belong to a much larger literature in cognitive science on the importance of causal structure in reinforcement learning \citep{gershman201717}.}

\paragraph{Sensitivity to prior information} Judgments of causal structure can be influenced in a graded manner by prior beliefs about the presence, strength, or functional form of causal links. \citet{griffiths2009theory} review various strands of evidence for this claim and demonstrate how judgments quantitatively match the implications of Bayesian structure learning. For example, manipulating the base rate of causation affects beliefs about whether objects cause a detector to activate, even with the same subsequent evidence. Other findings point to the role of informative priors over causal strength or functional form. In the classic task of elemental causal induction (see Section \ref{sec:elementalCausalInduction}), where people learn about a single cause-effect relationship, \citet{Lu2008GenericPriors} show that models with priors favoring sparse and strong causes explain both strength judgments and structure judgments better than models with uninformative priors. Likewise, \citet{Lucas2010HierBayes} demonstrate how priors over functional form influence judgments of structure by exposing participants to causal systems governed by either a conjunctive form (where both causes must be present to yield an effect) or a disjunctive form (where either alone suffices). Under a conjunctive form, repeated failures by a single candidate cause are not so informative, since it may need a partner cause; under a disjunctive form, the same failures count strongly against causation. Accordingly, participants given conjunctive training largely ignored such failures when later evaluating a novel candidate cause, as predicted by a Bayesian model.

\paragraph{Hierarchical Bayesian inference} \citet{Kemp2010LearnCausalModels} study whether people use what they learned in earlier, similar situations to form expectations in new ones. In their experiments, participants saw objects that affected a machine, where some objects had the same causal effects and could therefore be grouped together, even though these groupings were not directly stated. When asked to evaluate a new object, participants did not rely only on the limited data about that object. Instead, they interpreted this information in light of patterns they had seen before. As a result, the same sparse evidence (sometimes just visible features) led to different conclusions depending on what participants had previously learned. For example, after seeing that earlier objects fell into two groups with different effects, participants tended to assign a new object to one of these groups, so the same observation was interpreted differently depending on which grouping seemed more likely. Moreover, these expectations were flexible. When new observations repeatedly contradicted the earlier grouping, participants updated their beliefs and were willing to revise the grouping itself. A hierarchical Bayesian model captures this behavior by allowing people to update both which group an object likely belongs to and what effect objects in each group tend to have.

\subsubsection{Bounded rationality in structure learning}\label{sec:processesOfStructureLearning}

While the above evidence suggests that Bayesian structure learning provides a good as-if model of human causal cognition, it may be too computationally demanding to cognitively implement in full detail. Indeed, people rely on coarse, schematic representations that capture broad causal patterns while leaving many mechanistic details unspecified \citep{Keil2003Folkscience}. The \emph{illusion of explanatory depth} reveals that this incompleteness often goes unnoticed \citep{Rozenblit2002IOED}. More generally, people appear willing to tolerate or even prefer compressed causal models when additional detail seems unnecessary for the task at hand \citep{Kinney2024Compressed}.

How do humans actually learn causal structure? A prominent view in the cognitive science literature proposes that they infer larger causal structures through sequences of local updates rather than by globally evaluating all candidate structures. The evidence in \cite{Fernbach2009Local}, for instance, suggests that people  evaluate individual causal links in a piecemeal fashion, and focus on evidence that is immediately available and easy to process.  Specifically, when a variable $X$ is intervened on, people tend to treat the resulting pattern of activations on that trial as evidence for direct causal links from $X$ to each variable that becomes active, and they build toward a more global understanding by accumulating such local inferences. Consistent with this account, experimental participants made systematic errors when inferring chains $X\rightarrow Y \rightarrow Z$, frequently believing in a spurious, direct $X\rightarrow Z$ connection \citep[][document the same pattern]{Frechetteetal.}.

Purely local computations raise the concern that they need not result in a globally coherent belief system or correspond to a normatively well-grounded learning process. \cite{Bramley2017Neurath} addresses this issue by proposing a form of local learning based on Gibbs sampling, a method for approximate inference. They call this the `Neurath’s Ship' algorithm, alluding to a metaphor attributed to Otto van Neurath: `\textit{We [theorists] are like sailors who on the open sea must reconstruct their ship but are never able to start afresh from the bottom. Where a beam is taken away a new one must at once be put there, and for this the rest of the ship is used as support.}' In the algorithm, the learner maintains a single current DAG hypothesis and, after each new observation, performs a small number of local edits: it selects an edge and updates its state (absent, $i \rightarrow j$, or $j \rightarrow i$) by sampling conditional on the current graph and recent evidence. Updates are restricted to preserve acyclicity, and belief updating is both local and conservative, favoring small changes to the current hypothesis. In addition, past evidence is partially forgotten or downweighted, inducing a recency bias. This model matches two key empirical facts. First, people’s belief revisions about structure tend to be sticky; they rarely make large jumps between structures. Second, revisions exhibit a recency bias, consistent with the model’s assumption that older evidence is discounted.

Why does continued exposure to the DGP not necessarily correct misspecified beliefs? A number of papers speak to this issue, though not framed specifically in terms of DAGs. \cite{schwartzstein2014selective} shows theoretically that if agents do not initially attend to a variable, they may never learn its relevance and may instead become more convinced that the variables they do track are the important ones; \cite{hanna2014learning} observe this phenomenon among seaweed farmers. \citet{GagnonBartschetal.} extend this logic to characterize when agents with misspecified models will ever realize their model is wrong. A broad class of related \emph{learning traps} has been also studied in cognitive science \citep{RichGureckis2018}.

Even when the relevant models are known, people may be unable to fully compute their implications. In the experiment of \cite{MusolffZimmermann2025}, participants faced uncertainty about which of two explicitly specified models determined outcomes. The experimenters manipulated complexity by varying whether participants were given the already precomputed outcomes under each model or had to instead calculate the outcomes themselves before combining the models. Faced with the latter burden, participants calculated outcomes based only on the more likely model, despite otherwise continuing to acknowledge model uncertainty. In that sense, complexity can induce a shift from model averaging to model selection, not by changing which models people consider plausible but by making it harder to integrate them.

\section{Methodology: Inferring and eliciting beliefs about causality}\label{sec:measurement}

Most of the research discussed so far examines beliefs or judgments about specific causal relationships rather than complete belief systems. This raises two related questions. First, are individuals' choices or elicited beliefs consistent with a coherent subjective causal model? Evidence at the level of isolated comparative statics may appear consistent with the causal DAG framework even if judgments violate global structural restrictions (such as coherence between beliefs about direct, indirect, and total effects). Second, how can subjective causal structure be measured empirically, and to what extent do different elicitation methods capture the underlying belief system rather than artifacts of the measurement procedure?
The next two subsections tackle these questions in turn.

\subsection{Rationalizability}

\paragraph{Theory}
Theoretical work characterizes when a DM's choices can be represented as expected utility maximization under a single subjective causal model, and how that model can be identified. Approaches differ in the scope of feasible interventions as well as in the nature of the choice sets they consider and thus in the identifying variation they generate. \citet{HalpernPiermont} and \citet{Schenone} consider a setting in which the DM can freely intervene on any variable in the network. \citet{HalpernPiermont} recover a subjective DAG from preferences over interventions themselves (building on Pearl's machinery); \citet{Schenone} recovers them from intervention-act pairs, where interventions alter the subsequent evaluation of acts (building on Savage's machinery).

By contrast, \citet{EllisThysen} emphasize that economic agents typically control only a small subset of nodes in the causal system determining their utility, and that utility may in turn depend on few nodes. An employee, for instance, may only be able to control their effort and may just care about remuneration. Hence, they consider a DM who can affect merely a single node (the `action') and whose utility depends on a single node (the `outcome'), though their results extend to the case of multiple outcomes if these are all directly causally related to one another. This restriction makes identification more demanding, as the analyst just observes stochastic choices over actions rather than preferences over arbitrary interventions. Nonetheless, they show that if the utility function is known, the stochastic choice rule follows a logit form, and the subjective DAG contains no $v$-colliders, then choices from a single dataset identify the subjective DAG up to a relation they call \emph{prediction-equivalence}. Loosely, DAGs are prediction-equivalent if they share the same confounders (which they define as variables that the DM believes can influence both her action and another variable) and the same shortest directed paths from the action and each confounder to the outcome. The main contribution of \cite{EllisThysen} concerns the general case in which the utility function and the payoff node are unknown, and the data from which the DM learns may be affected by her own actions.

\paragraph{Are empirical beliefs DAG-rationalizable?}

The decision-theoretic work does not yet lend itself to empirical application because it either requires having subjects make an impractically large number of choices \citep{HalpernPiermont, Schenone} or makes the identifying assumption that individuals make no mistakes when fitting models to data \citep{EllisThysen}, in addition to the expected utility assumption that empirical choices generally violate.  
To empirically test whether individuals' beliefs are consistent with the causal DAG framework, \citet{AmbuehlHuang} devise the following method. They define a \emph{Causal Belief System} $\mathcal{B}$ as the list of all interventional probability statements defined over a set of random variables and elicit subjects' beliefs about all these probabilities. Letting $\mathcal{P}^D$ denote the corresponding interventional probability statements obtained by fitting DAG $D$ to the data, they call a subject's belief system \emph{DAG-mixture rationalizable} if there exists a probability distribution $q$ over all DAGs such that 
	$
	\mathcal{B} = \sum_{D} q_D \mathcal{P}^D
	$
and \emph{DAG rationalizable} if $q_D=1$ for some $D$. In other words, this test assesses whether a person's elicited beliefs could potentially stem from some weighted average of possible DAGs. Because the number of possible DAGs grows extremely quickly in the number of nodes, one might suspect that nearly any belief system is DAG-mixture rationalizable. However, this is not the case, because the predictions generated by different DAGs are highly constrained and overlap substantially, causing the vectors $\mathcal{P}^D$ to occupy a low-dimensional subspace of the space of all possible belief systems. The set of belief systems that can be represented as a mixture of DAGs is therefore much smaller than the set of all logically possible belief systems. Empirically, in a setting with relatively noisy DGPs, \cite{AmbuehlHuang} find that subjects' elicited belief systems are substantially closer to the best fitting DAG mixture than to a uniformly random benchmark---about three times closer in terms of Euclidian distance, both with and without allowing for probability weighting \citep{kahneman1979prospect}. At the same time, while the causal DAG framework has explanatory power, a substantial portion of variation remains unexplained.

\subsection{Elicitation}

Studying causal beliefs often requires measuring them directly. Existing methods to elicit causal structure fall into three broad classes: numbers (quantitative judgments), words (verbal descriptions), and pictures (graphical depictions). The approaches differ both conceptually and in how well they perform empirically. 

Conceptually, numerical elicitation has the advantage of imposing the weakest assumptions. Unlike graphical elicitation, it does not presume that human belief systems conform to the Causal DAGs formalism. Unlike verbal elicitation, it does not require the analyst to make decisions about the mapping from words to numerical quantities or DAGs. Yet, the elicitation of all quantities required to describe a causal network is large and quickly becomes unmanageable as the number of nodes grows. Moreover, requiring subjects to consider all possible relations also risks pointing out causal paths they might otherwise have ignored. 

Verbal elicitation has the advantage of most closely matching how people naturally talk about causation. Due to the unclear mapping between verbal statements and formal models, however, the resulting representations may be incomplete or ambiguous. Whether this incompleteness reflects an artifact of the elicitation method or a genuinely underspecified belief system remains an open question.

Graphical elicitation has the advantage of speed, completeness, and an ostensibly clear link to the formal framework. Conceptually, these advantages come at the price of assuming that individuals' causal belief systems are consistent with the causal DAGs formalism. It also needs modification to allow the expression of uncertainty over models. Even when these assumptions are satisfied, it is still unclear whether subjects interpret their own graphical depictions the same way an analyst would. If, for instance, a subject draws a collider DAG, does this mean that they are aware, and mean to convey, that conditioning on the collider will cause an apparent correlation between the source nodes? Moreover, the key information in a DAG is not the arrows that are present but those that are absent---arrows merely represent the possibility, but not the necessity of causal influence, whereas the absence of an arrow definitely precludes influence. It is unclear whether subjects appreciate this asymmetry when constructing diagrams.

Empirically, there is scant evidence that compares the approaches. One exception is \citet{AmbuehlHuang}, who elicit subjects' entire set of interventional and marginal beliefs over three binary variables after 100 rounds of interaction, and also ask them to depict their beliefs about causal relations graphically. They find that the list elicitation predicts choices well, but that the graphical depictions predict neither choices nor beliefs elicited in the list format. 

Other studies also compare list elicitation to graphical depictions, use benchmarks other than predicting consequential behavior. \citet{Tatlidiletal.} find---though under extremely strong identifying assumptions---that list elicitation produces elicited beliefs closer to the true DGP in a physical framing. They conjecture that this outperformance is due to the fact that list elicitation forces subjects to consider all potential causal connections. \citet{green2003eliciting} finds similar results in a social psychological context.

Verbal elicitation has been studied in a line of research that seeks to quantify the information conveyed by statements such as `A killed B' (strong) or `A caused B's death' (less strong) \citep{RoseSieversNichols2021, beller2025causation}, including through Large Language Models \citep{priniskiEtAlPipeline}. This literature leaves open several interesting questions about how verbally elicited beliefs relate to consequential choices, objective causal relations in the world, and alternative elicitation methods.

Within-study consistency is suggestive about the types of inferences that can and cannot be drawn from verbal statements.\footnote{A body of research has more generally studied the mapping between numerical probabilities and their verbal expression \citep{WallstenBudescu1983Encoding}.} \citet{Andre.InflationBeliefs} ask respondents \textit{`Which factors do you think caused the increase in the inflation rate? Please respond in full sentences.'} Two human research assistants then codify these responses in the form of DAGs. \citet{Andre.InflationBeliefs} find moderate inter-rater reliability on the level of individual links between nodes: if one coder assigns a causal connection between two specific factors, there is a 77\% chance that the other coder does so as well. Researchers seeking to understand subjective causal models, however, often need a more demanding benchmark than the coincidence of links, as it is three-node substructures that capture the key qualitative properties of causal DAGs. The rate at which the two coders map statements to exactly the same DAGs is 51\%. This number drops further once we consider DAGs that are not simply a list of factors which each influence the outcome (star-networks): if one coder maps the verbal statement to a DAG that involves at least one indirect effect, the chance that the other coder maps it to the same DAG is 21\%. The modest performance of the hand-labeling method is consistent with \citet{YangHanPoon2022} who also document its brittleness and report poor out-of-sample predictive performance. It is an open question whether this brittleness represents the ability of human research assistants (which natural language processing methods might solve), or inherent conceptual issues with the objective of mapping each respondent's verbal statement to a single DAG.\footnote{While \citet{Frechetteetal.} ask subjects to write notes to their future selves to help with prediction tasks in binary variable DAGs, they do not extract subjects' DAGs from these notes.}

These results illustrate that the choice of method must be informed by the researchers' aims. For instance, verbal elicitation may suffice if the goal is to identify which variables enter a subjective causal belief system, but more formal elicitation methods may be necessary to study beliefs about the structure of relations between these variables.

\section{Applications}\label{sec:applications}

Research on causal cognition has been used to inform political economy (Section \ref{sec:appPoliticalEconomy}), microeconomics (Section \ref{sec:appMicro}), macroeconomics and finance (Section \ref{sec:appMacro}), and business (Section \ref{sec:appBusiness}). Here, we briefly discuss a selection of research that applies the causal DAG framework in each of these fields.

\subsection{Political economy} \label{sec:appPoliticalEconomy}

Some of the most exciting applications of causal DAGs in economics concern political economy, where they provide a way to represent narratives as explicit causal models. This approach has been used to explore phenomena such as the persistence of competing narratives in public discourse, cycles of populism, and belief polarization.

The theory of narrative competition proposed by \citet{EliazSpiegler.Narratives} shows how conflicting causal narratives can persist simultaneously. It yields striking implications: false narratives can survive in equilibrium, and it is possible to predict the prevalence of narratives in a way that permits comparative statics. These results rest on the key assumption that individuals adopt whichever model promises the higher utility when fit to the data, without regard for its accuracy. To illustrate the logic, consider the debate over whether mask-wearing causally reduces COVID-19 transmission. Suppose there are two narratives, formalized as subjective causal models. Model 1 correctly states that increased masking reduces COVID-19 transmission. Model 2 holds that masking has no effect on transmission.  To see how, first consider the case in which most individuals adopt Model 1. In this case, most will wear masks, and case counts will be low. According to Model 1, this situation can be maintained only by continued masking, which has a hassle cost. Model 2 promises higher utility because it predicts case counts will remain low even if everyone stops masking and thus saves the hassle cost. Therefore, individuals will begin to adopt Model 2 and cease masking. Second, consider the case in which most individuals adopt Model 2. In this case, only few individuals will wear masks, and case counts will be high. According to Model 2, masking cannot change this situation. Model 1 promises higher utility because it predicts that case counts can be lowered at a small hassle cost. Individuals will thus flock to Model 1. Overall, therefore, the more popular one model is, the more attractive the other becomes; narratives `feed off each other.' Accordingly, no model, including the true one, can survive alone; narrative multiplicity is a necessary outcome. Equilibrium is achieved when the models are adopted with a frequency that makes them equally attractive. Comparative statics predictions arise from the fact that parameters such as the hassle costs of mask-wearing and the strength of causal effects in the DGP determine the equilibrium frequencies.

While \citet{EliazSpiegler.Narratives}'s model is insightful, its empirical accuracy remains contested. \citet{angrisani2024competing} demonstrate that it can rationalize the dynamics of Americans' beliefs about the effectiveness of preventive behaviors during the COVID-19 pandemic. Yet, \citet{AmbuehlThysen}'s direct experimental tests of its key assumption find little empirical support. The reason is that for a DM who views narrative adoption as a choice under ambiguity, \citet{EliazSpiegler.Narratives}'s key assumption is equivalent to maximizing best-case outcomes. Most of their subjects instead maximize worst-case outcomes, consistent with a general tendency observed in decisions under ambiguity \citep{ambiguityReview}.

Other work on subjective causal models in political economy does not assume that voters adopt narratives at will, but are instead married to a subjective causal model. Their turnout to the voting booth depends on the extent to which their model promises that their favored party would produce better outcomes than the competition. \citet{Levyetal.AER} show how this assumption, when placed in a dynamic political economy framework, can yield cycles of populism in which a populist party and a mainstream party take turns governing. Their mainstream party has the correct model of the world and chooses optimal policy, whereas the populist party's model omits some relevant variables. Turn-taking in governing arises due to the following process. When the populists observe outcomes generated under the mainstream party's policies, omitted-variable bias leads them to overestimate the effectiveness of the policies they consider relevant and to advocate more extreme policies. This raises their turnout and brings them to power. Once in power, their own policy outcomes gradually lead beliefs to correct, reducing their perceived advantage and eventually returning power to the mainstream party, at which point the cycle begins anew.

Also assuming that the strength of causal beliefs determines turnout, \citet{EliazGalpertiSpiegler} find that political coalitions will often find it optimal to spin false narratives that attribute desirable outcomes to the exclusion of particular social groups (`scapegoating') rather than the policies that were actually responsible. 

Other research uses the framework of causal DAGs to understand belief polarization, a phenomenon in which the same information leads people with different priors to update in opposite directions. Collider DAGs can naturally produce polarization through the logic of explaining away \citep[following the spirit of][]{botvinik2023belief}. Suppose a candidate's victory depends on two causes: actual majority support and election fraud. Consider two voters with different priors about majority support; one is confident that the majority favors the candidate, whereas the other is confident the majority opposes her. Both believe that fraud, if it occurs, would favor the candidate. Observing that the candidate won causes them to update their beliefs about fraud in opposite directions. The first voter, who expected the win based on majority support, sees fraud as less necessary to explain the outcome, while the second, who did not expect the win, finds fraud more plausible. Similar arguments can be constructed to provide rational explanations for many classical studies of belief polarization \citep{Jernetal.}.

\subsection{Microeconomics}\label{sec:appMicro}

\paragraph{Principal-agent theory} 

The causal DAG framework has yielded new insights in principal-agent theory in settings such as contracting and persuasion.

\citet{SchumacherThysen} study contracting when the agent evaluates actions through a misspecified causal model. Their main example features a marketer who seeks to maximize sales, and must decide whether to make cold calls. In their setting, cold calling makes customers more informed but damages the firm’s reputation, and both of these factors in turn affect sales. However, the agent's causal DAG is misspecified: she neglects the reputation channel and thus thinks cold calling pays off more than it really does. The principal can therefore induce an action that would otherwise be too costly, because the agent overestimates its effectiveness.

\begin{figure}[t]
\caption{Causal persuasion \citep{BurkovskayaStarkov2026CausalPersuasion}.}
\label{fig:causal_persuasion}
\centering
\includegraphics[width=0.9\textwidth]{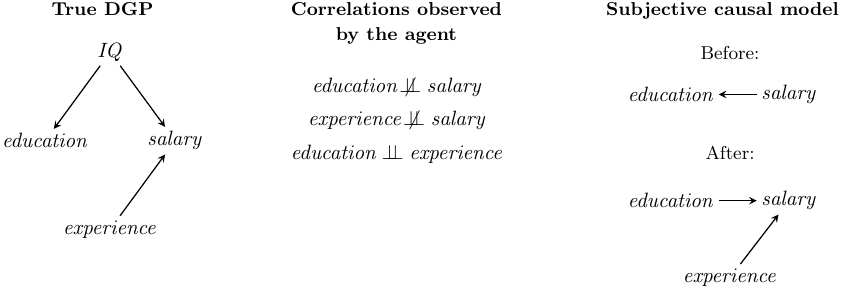}
\end{figure}

A larger number of papers study persuasion. Some of the most active research on persuasion through suggested interpretations relies on the \emph{model persuasion} framework by \citet{SchwartzsteinSunderam} that conceives of subjective models as likelihood functions but does not explicitly involve causality. In their baseline model, a principal and an agent both observe the same state-dependent dataset. The principal would like to induce an action that differs from the agent's preferred choice in that state. She does so by proposing an interpretation of the data in the form of a likelihood function, which the agent adopts if it fits the data better than his default model. Two central results are that, with finite datasets, persuasion is generally possible, and that agents tend to follow overfitted models.\footnote{Related research extends this framework in several directions: to organizational settings \citep{SchwartzsteinSunderam.Shared}; to environments in which the principal does not observe the specific data realization before deciding which model to send \citep{ChiaraAina}; to settings that incorporate elements of Bayesian Persuasion \citep{IchihashiMeng, jain2023}; and to experimental tests of whether subjects select the best-fitting model \citep{aina2025weighting}.} \citet{BarronFries} test this model experimentally and show that human subjects indeed manage to influence other's actions by proposing interpretations even when both observe the same data. Further expanding this line of research using the toolbox of causal DAGs is promising for two reasons. First, whenever there are multiple variables, likelihood functions are impractical to convey and may be replaced with causal explanations that can be modeled as causal DAGs. Second, assessing the quantitative fit of any model is difficult for humans. But causal stories in the form of DAGs allow subjects to evaluate models based on qualitative fit (e.g., whether the data display the presence or absence of correlations implied by the story), which many individuals manage to do well \citep{AmbuehlThysen}.

\citet{BurkovskayaStarkov2026CausalPersuasion} take a step in this direction. They analyze a setting in which a principal who knows the DGP can influence the beliefs of an agent who is unaware of some variables. The agent interprets data by forming subjective causal DAGs in the manner of a constraint-based algorithm (see Section \ref{sec:structureLearningConcepts}). They show that making the agent aware of neglected variables can lead him to reverse his beliefs about the direction of causality between other variables. Their main example, illustrated in Figure \ref{fig:causal_persuasion}, centers on a worker who believes that $education \gets salary$ when in fact both $education$ and $salary$ are independent effects of a third variable $IQ$ of which the worker is unaware. The agent is also unaware of the variable $experience$ that influences $salary$. The principal seeks to reverse the agent's beliefs about the direction of causation between $education$ and $salary$. She achieves this by pointing out that \emph{experience} is correlated with \emph{salary} but independent of \emph{education}. The worker infers that the only DAG consistent with this pattern of correlations is $education \to salary \gets experience$. \citet{BurkovskayaStarkov2026CausalPersuasion}'s general analysis of such persuasion strategies finds that persuasion is only possible when the receiver’s initial model conflicts with the true causal structure, and that making an agent believe in the existence of a link is often easier (in the sense that it requires disclosing just one or two variables) than making them believe that a link is absent (which requires disclosing every common cause). It is an open empirical question whether disclosing correlations between neglected variables causes agents to reassess the direction of causality in the way the model assumes, or instead prompts them to search for additional variables that reconcile the newly revealed correlations with their initial beliefs \citep{papakonstantinou2024humans}.

\paragraph{Behavioral economics} 

Many applications of the causal DAG framework intersect with behavioral economics, which frequently studies the impact of biased beliefs and imperfect inference. We have already covered formal connections to other theories in behavioral economics (see Section \ref{sec:misspecifiedLearning}) and contributions to topic areas that typically fall within its purview, such as self-confirming false beliefs (section \ref{sec:misspecifiedLearning}), stereotyping (section \ref{sec:parameterLearningEmpirics}), and failures to detect misspecification (section \ref{sec:processesOfStructureLearning}). Causal DAGs also bear on other issues of relevance, including moral decision making \citep[through the attribution of blame and credit;][]{lagnado2013causal, gerstenberg2018lucky, engl2022theory}, categorical thinking \citep{rehder2001causal, rehder2017categorization, rehder2017concepts}, and confirmation bias and motivated reasoning \citep{Gershman2019NeverWrong, Caddick2021Motivated, pilgrim2024confirmation}.

\subsection{Macroeconomics and finance} \label{sec:appMacro}

Macroeconomics is a fruitful field for applying the causal DAG framework, since expectation formation plays a central role in many macroeconomic questions. Applications have taken both theoretical and empirical directions. 

Theoretically, \citet{Spiegler2020} revisits the classical question of whether the Phillips curve can be systematically exploited by a central bank, but under the assumption that the private sector forms beliefs by fitting a a DAG to observed data, rather than holding rational expectations. In his leading example, the private sector endorses a `classical' narrative, wherein it perceives output as a cause of inflation rather than the reverse, and excludes expectations from its model entirely, as did monetary theory prior to \citet{phelps1967phillips} and \citet{friedman1968role}. Because this DAG treats the central bank's action and output as independent co-causes of inflation without linking them directly, it contains a $v$-collider. In this case the central bank can systematically push inflation above private-sector expectations. However, when the private sector's DAG is perfect (that is, when it does not contain a $v$-collider), or when the objective distribution is multivariate normal, no such exploitation is possible (see Section \ref{sec:misspecifiedLearning}). This result puts the historical debate surrounding the Lucas critique into a new perspective: non-exploitability results do not necessitate rational expectations; they survive under a much broader class of causal models.

While \citet{Spiegler2020} asks whether a strategic outside party can exploit an agent's causal misperceptions, \citet{Spiegler.MonetaryPolicy} shifts the focus to what happens when the agent with the wrong causal model is the DM, so that her behavior shapes the very data from which she draws faulty inferences. The leading macroeconomic application is again the Phillips curve. Following \citet{sargent1999conquest} and \citet{chokasa2015}, the private sector's subjective model inverts the Phillips curve, meaning that it treats employment as an independent explanatory variable for inflation, whereas in the true model causality runs from inflation to employment. The true DGP also includes fundamentals that directly affect inflation but not employment. When the private sector bases its inflation forecasts on this inverted Phillips curve, forecast errors arise when anticipated inflation has real effects, as follows.\footnote{Under the new-classical assumption that only unanticipated inflation matters, no systematic forecast errors arise. The reason is that in this case, employment reduces to the forecast error plus noise, which is by construction uncorrelated with the predictors in the private sector's inverted Phillips curve; treating it as exogenous is therefore harmless.} To forecast inflation, the private sector conditions on fundamentals and employment and then averages over employment using its unconditional distribution, as if employment were an exogenous shock unrelated to fundamentals. But when anticipated inflation has real effects, employment is in fact correlated with fundamentals through its dependence on inflation. By averaging over employment as if this correlation did not exist, the private sector dilutes the information that fundamentals carry about inflation, pulling forecasts toward the unconditional mean and making them too rigid. The extent of this dilution depends on the private sector's own forecasting behavior. A more rigid forecast entails less variation in anticipated inflation, which changes how much employment covaries with fundamentals and thereby how much the exogeneity assumption distorts the next round of inference. The personal equilibrium is the forecasting rule at which the rigidity implied by this distortion exactly matches the rigidity that generated the data to which the wrong model was fit. Hence, in equilibrium, forecasts underrespond to changes in fundamentals.

Empirical work on causal cognition in macroeconomics shows how misspecified beliefs about the causal structure of the macroeconomy cause demand for counterproductive policies. \citet{binetti2024people}, for instance, find that 60\% of Americans erroneously believe that high interest rates cause high inflation and support rate cuts to fight it. This finding parallels \citet{Andreetal.MacroNarratives}, who show more broadly that laypeople's intuitions about macroeconomic mechanisms differ systematically from those of experts. Regarding the relation between interest rates and inflation, they find, as do \citet{binetti2024people}, that a majority of households predict that an interest rate hike increases inflation whereas most experts predict the opposite. They show that households disproportionately think of a `cost channel' in which firms pass higher borrowing costs on as higher prices, whereas experts invoke the textbook demand channel. While theoretical models often employ the representative agent paradigm, \citet{Andreetal.MacroNarratives} also document substantial heterogeneity in households' subjective models of the macroeconomy. Relatedly, \citet{Andre.InflationBeliefs} study narratives about the 2021-2022 U.S. inflation surge and find that households' narratives are simpler and more fragmented than those of experts and differ systematically in terms of the variables they consider to be causes of inflation. Households focus disproportionately on supply-side factors such as supply-chain disruptions and the energy crisis, and often feature politically loaded explanation such as government incompetence or corporate greed which are entirely absent from expert narratives.

The implications of misspecified subjective mental models have also been studied in finance, partly following the influential presidential address of \citet{Shiller.Narratives}. While this work does not explicitly deploy the causal DAG framework, research by \citet{Molavietal.}, for instance, is closely related. They consider an asset-pricing environment in which fundamentals are driven by a hidden $n$-factor model. Agents can observe the realized fundamentals but not the underlying factors. Agents seek to understand these fundamental processes by estimating hidden-factor models. Due to complexity constraints, these models feature fewer factors than are present in the DGP. This model generates both predictable returns and predictable forecast errors. When applied to foreign exchange markets, it produces two classic violations of uncovered interest rate parity, namely the forward discount puzzle and the predictability reversal puzzle.

\subsection{Business} \label{sec:appBusiness}

Research on causal cognition has furthered our understanding of both marketing and strategic management by clarifying how consumers and managers perceive causal relationships and act on those perceptions.

\paragraph{Marketing}

One stream of research in marketing investigates which kinds of claims about a product's causal mechanism are convincing to consumers. 
\citet{fernbach2013explanation} find that more reflective individuals (as measured by the cognitive reflection test) prefer more detailed mechanistic explanations of how products work.
\citet{BhartiSussman2025DirectionalChains} examine product claims that invoke causal chains, and show that people prefer products when all links in that chain are directionally consistent (e.g., a supplement increases alertness by increasing hormone production) compared to when the chain involves both positive and negative links (e.g., the supplement increases alertness by reducing hormone production). They argue that directionally consistent chains are easier to process, in turn raising perceptions of product efficacy.

\citet{KuporLaurin2020ProbableCause} and \citet{DanielsKupor2023MagnitudeHeuristic} study the link between the perceived probability and magnitude of product outcomes. The former paper shows that people expect more likely outcomes to be larger; for example, when participants were told that tomato consumption had a higher probability of increasing metabolism, they expected the resulting increase in metabolism to be greater. The latter paper documents the converse pattern, wherein people infer that relationships with larger perceived magnitudes are more likely to reflect causal effects. Both effects are consistent with the interpretation that people believe that strong underlying causes increase both the probability and the magnitude of effects.

Other research examines \emph{causal dilution}, the sometimes-observed phenomenon that the perceived strength of a cause declines with the number of effects it has. \citet{SussmanOppenheimer2020Effectiveness} found causal dilution when additional effects were beneficial---subjects judged products to be less effective for their main purpose---but the opposite when the effects were harmful. \citet{stephan2023perceived} suggest that the lack of dilution in the latter case may have resulted from a kind of causal elaboration in which participants spontaneously added causal links among the negative outcomes. For example, if a shaving cream causes both ingrown hairs and skin irritation, people may infer that the ingrown hairs also worsen the irritation. Controlling for this mechanism yielded causal dilution in both positive and negative conditions. \citet{zhang2026more} point out that when causal structure is uncertain, the opposite of causal dilution can occur. Accordingly, experimental participants inferred more strongly that drinking tea improves bone health when they also learned that tea consumption is correlated with better heart health. The authors argue that when people simultaneously evaluate both structure and strength, additional correlations make the focal outcome seem more plausibly affected by the same cause.

Causal DAGs can shed light other phenomena in marketing such as \emph{backfire effects} \citep{bhui2020paradoxical}. These arise when information that appears favorable is interpreted negatively. A product's star rating in online retail, for instance, could reflect both genuine quality and fake reviews. Under suitable parametric assumptions, Bayesian observers treat ratings that are high but still plausible as evidence of product quality, but interpret implausibly high ratings as a sign of fake. Inferred product quality can thus decline when signals are considered too good to be true. Experimental participants in such a setting became more likely to suspect fraud as ratings increased, especially when prior information made fraud seem more likely. Stronger fraud inferences were associated with a weaker or even negative effect on purchase attitudes, and in many cases the same individuals responded both positively and negatively to higher ratings, depending on that prior information.

\paragraph{Strategic management}

A research program in strategy termed \emph{Bayesian Entrepreneurship} \citep{Agrawal2026BayesianEntrepreneurship} draws on insights from Bayesian learning to understand and improve entrepreneurship. One of its primary tenets emphasizes the heterogeneous priors that individuals hold. DAGs afford us a means of characterizing this heterogeneity beyond simple quantitative differences in degrees of belief, capturing instead more fundamental differences in how individuals see the world. \citet{camuffo2024theory}, for instance, argue that entrepreneurs' novel theories---formalized in their framework as DAGs---are a key source of competitive advantage. They illustrate this hypothesis with the case of the eyewear manufacturer Luxottica, which rose to prominence by reconceptualizing eyewear as a part of the fashion industry.
This work falls within a broader literature on the \emph{theory-based view of strategy} \citep{FelinZenger}, with \citet{ehrig2024causal} focusing specifically on causal reasoning in the tradition of \citet{Pearl.2009}. Empirically, \cite{camuffo2024scientific} show in a series of large randomized controlled trials that this theory-driven, evidence-based approach \citep[e.g.,][]{felin2021value} improves entrepreneurial decision making.

\section{Conclusion and further reading}\label{sec:conclusion}

The history of economics features many examples of mathematical frameworks that had been fruitfully applied in other sciences for decades before economists started using them, including calculus, the theory of random networks, and information theory. We believe that causal networks offer another such valuable addition to the economist's toolkit.

Given its multi-decade interdisciplinary history, it is impossible to cover all aspects of causal cognition in an article-length review. Among the many fascinating questions that exceed the scope of this review are the following. 

What are the entities that are linked by causal networks? This is a question of \emph{causal ontology} \citep[see][]{griffiths2009theory}. It is relevant, for example, to differences in economic intuition between experts (who think about GDP, unemployment, and the money supply as separate entities) and laypeople (who may collapse these into a single entity called `the economy'). Related work in cognitive science examines how individuals identify what belongs in a causal system and how their concepts change with experience \citep{Goodman2007Grounded, rottman2012causal, goldwater2015acquisition}.

Where should the boundaries of a causal network be drawn, given that in principle every variable in the universe may be related to every other? 
This is the \emph{causal frame problem}  \citep{Shanahan2004FrameProblem}. One approach in cognitive science is to treat it as a cost-benefit analysis where the greater accuracy from larger and more detailed models is weighed against their computational cost \citep{icard2015resource}. Of course, this cannot be a complete solution since it still presupposes a way of figuring out which parts of a potentially unbounded causal model are worth considering in the first place.

What leads people to single out some causes rather than others?
Why, for instance, do we call an arsonist the cause of a fire but not the atmosphere, even though both the lit match and the presence of oxygen were necessary for the house to burn? This is the domain of research on \emph{causal responsibility} judgments \citep[e.g.,][]{lagnado2013causal}, which shows that people hold remarkably consistent intuitions as to which of multiple events in a causal model is the most important cause of an outcome \citep{morris2018judgments}. The economic model by \citet{AlexanderGilboa} hypothesizes that people regard $x$ as a cause of $y$ to the extent that accounting for $x$ reduces $y$'s Kolmogorov complexity. In economic policy applications, one might conjecture that such a systematic focus may direct attention toward specific policies (such as attributing economic development to new road construction) while neglecting the necessity of others (such as maintaining water infrastructure).

How do people make inferences when we cannot observe all variables in a causal network? Reasoning about such \emph{hidden causes} has been studied extensively in cognitive science  \citep[e.g.,][]{hagmayer2007inferences, rottman2011unobserved, Kushnir2010Hidden}. In economics, \citet{samuelson2025robust} study this issue within a broader framework based on variational autoencoders. 

How do people draw parallels between different causal systems? We have already seen Bayesian models which incorporate forms of abstraction permitting agents to generalize their knowledge across settings. Generalization can extend further still as exemplified by analogical reasoning, where systems may be similar only in their relational structure even though the variables themselves are entirely different \citep{holyoak2010analogical}. We speculate that future research in this vein might be informative about lay economic reasoning. For example, insights from applications of causal DAGs to intuitive physics \citep[reviewed in][]{gerstenberg2017intuitive} could shed light on how individuals (mis)apply concepts such as momentum or velocity in their mental models of economic phenomena.

The causal DAG framework has also been applied to illuminate fields outside of economics and cognitive science. In legal scholarship \citep[reviewed in][]{lagnado2017causation} the framework facilitates progress on difficult conceptual questions about how to formalize notions like the \emph{probability of necessity and sufficiency} \citep{Pearl.2009} that clarify cases such as the arsonist example above. Pertaining to public health, \citet{powell2023modeling} show that a detailed understanding of people's causal theories surrounding vaccines can help with the development of more effective informational interventions. Recent research also applies paradigms from cognitive science to artificial intelligence, in tests of whether Large Language Models reason causally like humans \citep{DettkiLakeWuRehder2025}, representing a return to the framework's computational roots.

From Aristotle to AI, the study of causation traces a venerable intellectual history to which economists are no strangers. Yet economics has only begun to engage with the flourishing literature on how ordinary people think about cause and effect, one that currently links fields as diverse as psychology, philosophy, and computer science. The research reviewed in these pages offers a sophisticated suite of theoretical and empirical tools that mesh well with existing approaches. We see myriad opportunities to understand how people construct their worldviews, to explore what leads them to diverge, and ultimately to close the gap between economic theories and economic life. The returns to this enterprise are, in our mental models, substantial.

\newpage

\setstretch{1}
\setlength{\bibsep}{0pt plus 0.3ex}
\bibliographystyle{chicago}
\bibliography{narratives}

\end{document}